\pdfoutput=1 

\documentclass[3p]{elsarticle}

\biboptions{sort&compress}
\usepackage{lineno,hyperref}
\usepackage{caption}
\usepackage{graphicx}
\usepackage{subfigure}
\graphicspath{{figures/}}
\modulolinenumbers[5]

\usepackage[namelimits]{amsmath}
\usepackage{amssymb}

\usepackage{booktabs}
\usepackage{multirow}
\usepackage{array}

\journal{Journal of \LaTeX\ Templates}

\begin{document}

\begin{frontmatter}

\title{Closed loop image aided optimization for cold spray process based on molecular dynamics}

\author[mymainaddress,mysecondaryaddress]{Zhenxing Cheng}
\author[mymainaddress]{Hu Wang \corref{mycorrespondingauthor}}
\cortext[mycorrespondingauthor]{Corresponding author}
\ead{wanghu@hnu.edu.cn}
\author[mysecondaryaddress]{Gui-rong Liu}

\address[mymainaddress]{State Key Laboratory of Advanced Design and Manufacturing for Vehicle Body, Hunan University, Changsha, 410082, PR China}
\address[mysecondaryaddress]{Department of Aerospace Engineering and Engineering Mechanics, University of Cincinnati, Cincinnati, Ohio, 45221, United States}

\begin{abstract}
This study proposed a closed loop image aided optimization (CLIAO) method to improve the quality of deposition during the cold spray process. Some recent research shows that the quality of deposition measured by flattening ratio of the bonded particle is associated with impact velocity, angle and particle size. Therefore, the original idea of CLIAO is to improve the quality of deposition by obtaining the maximum flattening ratio which is extracted from the molecular dynamics (MD) simulation snapshots directly. To complete this strategy, a Python script is suggested to generate the required snapshots from result files automatically and the image processing technique is used to evaluate the flattening ratio from the snapshots. Moreover, three optimization methods including surrogate optimization (Efficient Global Optimization) and heuristic algorithms (Particle Swarm Optimization, Different Evolution algorithm) are engaged. Then a back propagation neural network (BPNN) is used to accelerate the process of  optimization, where the BPNN is used to build the meta-model instead of the forward calculation. The optimization result demonstrates that all the above methods can obtain the acceptable solution. The comparison between those methods is also given and the selection of them should be determined by the trade-off between efficiency and accuracy.

\end{abstract}


\end{frontmatter}

%\linenumbers

\section*{Highlights}
	\begin{itemize}
		\item A closed loop image aided optimization is suggested for the cold spray process;
		\item Several optimization methods are engaged including surrogate model and heuristic algorithms;
		\item An image processing technique is used to evaluate the objective function from the result contours directly; 
		\item A back propagation neural network is constructed to accelerate the process of classic optimization methods.
	\end{itemize}

\section{Introduction}

\begin{figure}[htbp]
	\centering
	\includegraphics[width=8cm]{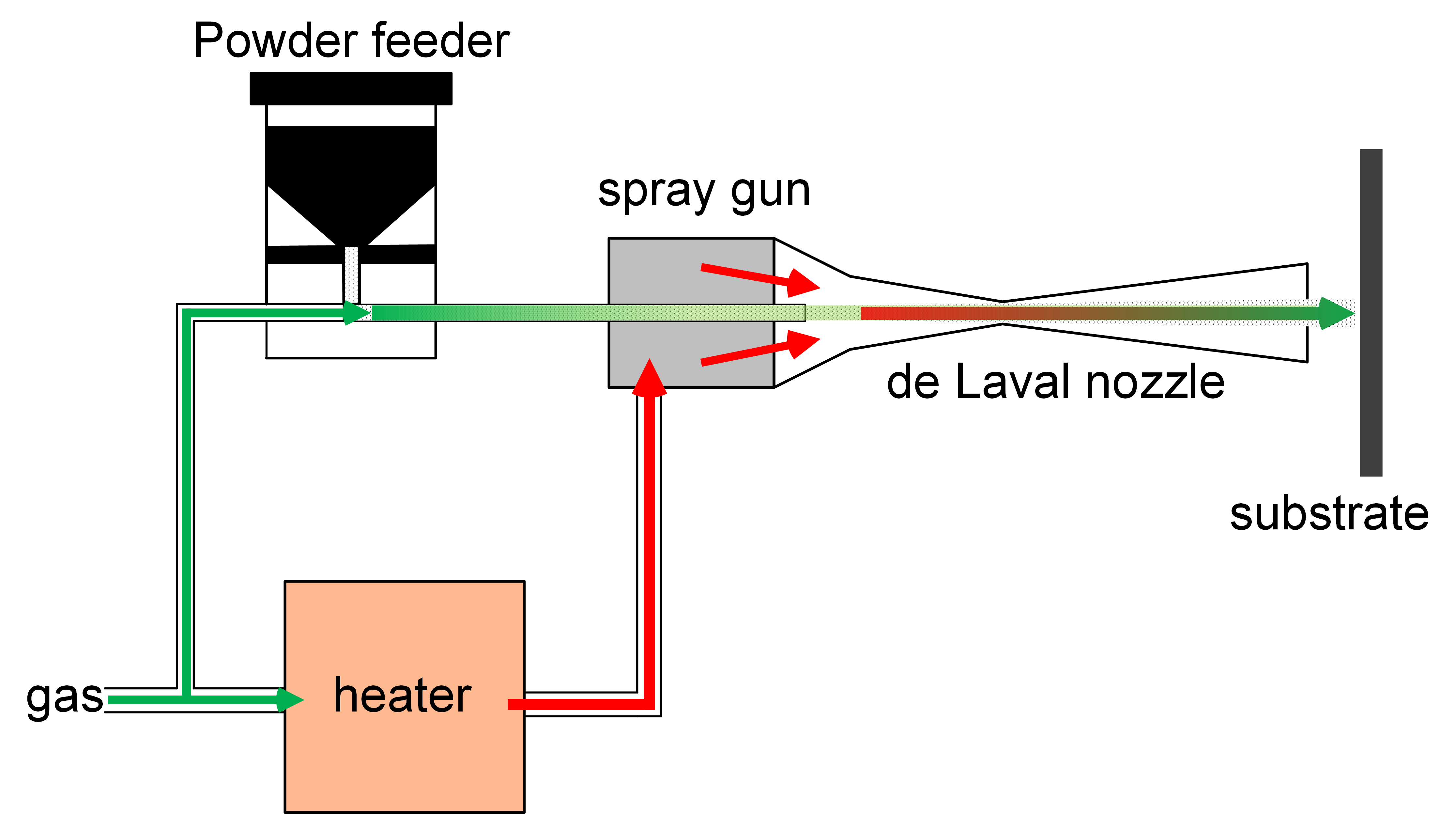}
	\caption{Schematic of Cold Spray Process \cite{assadi2003bonding}.}
	\label{fig:schematic}
\end{figure}

Cold spray is an innovative solid-state material deposition process, where micron sized particles bond to the substrate as a result of high-velocity impact. Compared with thermal spay, the powder remains in the solid state during the entire process which results in reduced oxidation and absence of phase transition. Therefore, cold spray is widely used to temperature sensitive (such as nanostructured and amorphous) and oxygen sensitive (such as aluminum, copper, titanium, zinc, etc.) materials \cite{Coatings2017}. During the cold spray process, the particles are accelerated to the speeds between 300-1200 m/s by propellant gas and then impact on the substrate to form the plastic deformation and bonding connection which is shown in Figure \ref{fig:schematic}. Cold spray was originally developed as a coating technology in the 1980s \cite{alkhimov1990method} and then became one of the most popular solid-state process due to its distinctive features in repairing turbine and compressor blades without changing their highly complex crystal structure \cite{AdditiveManufacturing2018}. Along the way, a number of academic studies on cold spray have been proposed which covering various aspects of the process and its applications \cite{ActaM2016,JMS2015,papyrin2006cold,JMST2018,AdditiveManufacturing2018}. Generally, those studies can be divided into two categories: mathematical modeling of impact using numerical methods and experimental investigations. 

As for the mathematical modeling, there are numerous studies on numerical simulation of cold spray process. Finite element method (FEM) is one of the wide-used numerical methods for cold spray and the finite element method software ABAQUS Explicit has been widely used to simulate the particle deformation \cite{jtst2013,jtst2011,jtst2009,ass2009}, bonding features and associated mechanisms \cite{actam2008,actam2009,SCT2016influ} in cold spraying. These studies are based on Lagrangian formulation. A general feature in Lagrangian-based FEM simulations is that they can demonstrate the interface between particle and substrate during the process of high speed impact, but it cannot reveal the realistic development of the deformation pattern due to the excessive mesh distortion. Nevertheless, the Eulerian-based FEM does not have such drawback, so it offers a more realistic account of the deformation geometry \cite{ActaM2016}.Then an Eulerian-based FEM software package named CTH, which was developed by Sandia National Laboratories, has been used to analyze the interfacing bonding in the cold spray process \cite{ass2003computa}. Furthermore, both Lagrangian and Eulerian FEMs are mesh-based methods. There have also been some attempts to simulate cold spray using mesh-free methods, such as smoothed particle hydrodynamics (SPH) and molecular dynamics (MD). SPH is a non-mesh-based numerical method that can avoid the problems associated with extreme mesh distortion. Studies of simulating cold spray using SPH include the work by Manap et al. \cite{se2014experi} and Li et al. \cite{ass2010numerical}. However, the process of bonding and rebounding is still cannot be demonstrated  by neither FEM nor SPH method because the bonding happens at atom scale during the cold spray process. Therefore, MD method is considered as an ideal tool to simulate the bonding of cold spray but there are few studies focused on this. The earlier research is relevant to the aerosol deposition which revealed that higher impact velocities led to a stronger interface \cite{jtst2013impact,cms2014atomistic}. Recently, some scholars and researchers have made progress in simulating cold spray process by MD method. For instance, Yao et al. have simulated collision behavior between nano-scale TiO{\tiny 2} particles during cold spray \cite{jnn2018md} and Joshi et al. modeled the whole cold spray process in nanometer dimension \cite{jmp2018molecular}. These studies investigated the bonding mechanism in cold spray process and understood the effect of critical parameters including impact velocity, angle and particle size.

According to the Reference \cite{jmp2018molecular}, it is evident that the quality of deposition is associated with impact velocity, angle and particle size, where the quality of deposition is evaluated by measuring the height of deposition and flattening ratio of the bonded particle. It can be found that different impact velocities, angles and particle sizes result in different flattening ratios, so how to obtain the optimal flattening ratio by the corresponding parameters (velocity, angle, particle size)? This study proposed a closed loop image aided optimization (CLIAO) method to improve the quality of deposition by obtaining an optimal flattening ratio. In order to form a closed loop optimization, the image processing technique is used to calculate the flattening ratio from the result contours directly. Moreover, several popular optimization methods are employed to obtain the optimal solution which include efficient global optimization (EGO) \cite{jones1998efficient}, particle swarm optimization (PSO) \cite{kennedy2011particle}, differential evolution (DE) algorithm \cite{storn1997differential}. Furthermore, the back propagation neural network (BPNN) is used to accelerate the optimization process by building a meta-model for the objective function.

The rest of this paper is organized as follows. The theory of the proposed method is introduced in section 2. The results are shown in section 3 and the discussions should be given too. At the last, the conclusion is summarized in section 4.

\section{Methods}

\subsection{Framework of CLIAO method}

\begin{figure}[htb]
	\centering
	\includegraphics[width=12cm]{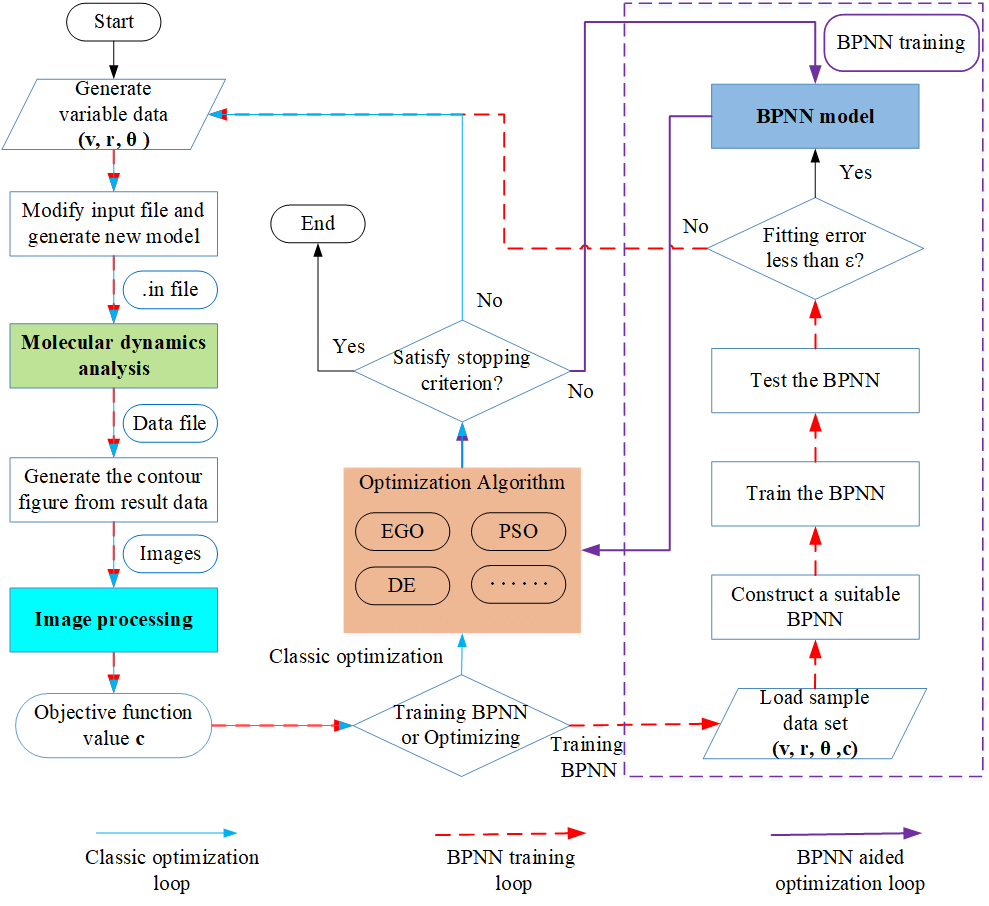}
	\caption{Framework of the CLIAO method}
	\label{fig:framework}
\end{figure}

The CLIAO method is suggested to obtain the optimal flattening ratio during the cold spray process. The key point is how to calculate the flattening ratio from the result image files instead of data files. Figure\ref{fig:framework} is the framework of the CLIAO method. It is visible that the CLIAO method contains three loops: the classic optimization loop, the BPNN training loop and the BPNN assisted optimization loop. Obviously, the classic optimization loop is a general process of most methods. Firstly, calculate the value of the objective function, then update the design variables, then calculate the new value of the objective function, then repeat this cycle until satisfying the stopping criterion. MD and image processing technique are used to calculate the value of the objective function in this loop. The second loop, BPNN training loop, is used to build a mate-model of the objective function. Once the BPNN model has been constructed, the value can be evaluated from the BPNN model instead of forward calculation. It can extremely improve the efficiency of classic optimization. The third loop is the BPNN aided optimization loop. Compared with the classic optimization loop, the only difference is the value of objective function should be obtained from the BPNN model directly. More details can be found in the next sections.

\subsection{MD simulation for cold spray}

\subsubsection{Basic theories of MD}

In the MD simulation, the system is described by the position and momentum of each atom or molecule in the simulation box. The dynamics of atoms obey Newton's law which can be described as the following equation:

\begin{equation} \label{eq:1}
	\begin{split} 
		&\dot{x}_i = v_i,\\
		&m_i \dot{v}_i = -\triangledown_{x_i}V,
	\end{split}
\end{equation}

\noindent where $m_i$ and $v_i$ are the mass and velocity of atom $i$ respectively, $x_i$ means the position of atom $i$ and $V(x_1,x_2,x_3,\ldots,x_n)$ denotes the inter-atomic potential. A typical example of the inter-atomic potential is the Embedded Atom Method (EAM) in which an atom should be regarded as the embedded component of a lattice \cite{daw1983semiempirical}. Specially, the EAM potential can be written as

\begin{gather}
	U = \sum_i F_i\left( \rho_i \right)  + \frac{1}{2}\sum_i \sum_{j\neq i} \phi_{ij}\left( r_{ij}\right), \\
	\rho_i = \sum_{j\neq i} \rho_j \left( r_{ij}\right) , \label{eq:3}
\end{gather}

\noindent where $F_i$ means the embedding energy depended on the electron cloud density $\rho_i$ around the atom $i$. The electron density $\rho_i$ is associated with all the atoms in the system which can be calculated by Equation\eqref{eq:3}. The symbol $\phi_{ij}$ denotes the pairwise potential, which depends on the relative distance $r_{ij}$ between atom $i$ and its neighbor atom $j$. Generally, most empirical potentials can be written as

\begin{equation}
	V=\sum_i V_i (u_1,u_2,u_3,\ldots,u_n),
\end{equation}

\noindent where $V$ is a function of the energy of each atom ($V_i$), which depends on $u_i$, the displacement of atom $i$ from its reference position $R_i$ ($u_i=x_i-R_i$). Usually, atoms do not interact directly beyond the cut-off radius $r_{cut}$, which implies that

\begin{equation}
	\triangledown_{x_i}V_i = 0,   if r_{ij}>r_{cut}.
\end{equation}

Therefore, the inter-atomic  force $f_i$ on the atom $i$ can be witten as

\begin{equation}
	f_i=-\triangledown_{u_i}V = \sum_{j \neq i}f_{ij}.
\end{equation}

There are a series of motion equations needs to solving during the MD simulation, such as Equation\eqref{eq:1}. Thus, the Velocity-Verlet algorithm is employed as the time integration algorithm to solve the motion equations with a considerable accuracy. More details about the Velocity-Verlet algorithm can be found from the Reference\cite{omelyan2002optimized}.

\subsubsection{Definition of stress}

The definition of stress for an atomic simulation is different from the continuum stress concept. A well-known definition of virial stress suggested by Swenson et al. \cite{swenson1983comments} is used in this study. Atomic scale virial stresses are equivalent to the continuum Cauchy stresses \cite{subramaniyan2008continuum}. The stress contains two parts, potential and kinetic energy parts, which is defined as

\begin{equation}
	\sigma_{xy}=\frac{1}{V}\sum_i \left[ \frac{1}{2}\sum_{j = 1}^N \left( r_x^j - r_x^i \right) f_y^{ij}-m^i v_x^i v_y^i \right] ,
\end{equation}

\noindent where $m^i$ means the mass of atom $i$, the subscripts $x$ and $y$ denote the Cartesian components and $V$ is the total volume of the system. The superscripts $i$ and $j$ are the atom identification number, which mean atom $i$ and atom $j$. The symbols $r$, $f$ and $v$ indicate the relative position, inter-atomic force and velocity respectively. Specially, the symbol $f_y^{ij}$ is the $y$ direction force on atom $i$ induced by atom $j$, $v_x^i$, $r_x^i$ are the velocities and relative position of atom $i$ along the $x$ direction. Other symbols are as similar as the above. To roughly calculate the local stress field of the system, the 'atomic stress $\sigma_{xy}$' for each atom in the system is used to plot the stress contours. Here, the 'atomic stress $\sigma_{xy}$' has the unit of stress$\times$volume. Then the 'Von Mises stress $\bar{\sigma}$' can be calculated by

\begin{equation}\label{eq:8}
	\bar{\sigma}=\frac{1}{\sqrt{2}}\sqrt{(\sigma_x-\sigma_y)^2+(\sigma_y-\sigma_z)^2+(\sigma_z-\sigma_x)^2+6(\tau_{xy}^2+\tau_{yz}^2+\tau_{zx}^2)},
\end{equation}

\noindent where $\sigma_x, \sigma_y, \sigma_z$ are the normal stresses and $\tau_{xy}, \tau_{yz}, \tau_{zx}$ are the tangential stresses.

\subsubsection{MD model for cold spray}

\begin{figure}[htbp]
	\centering
	\includegraphics[width=12cm]{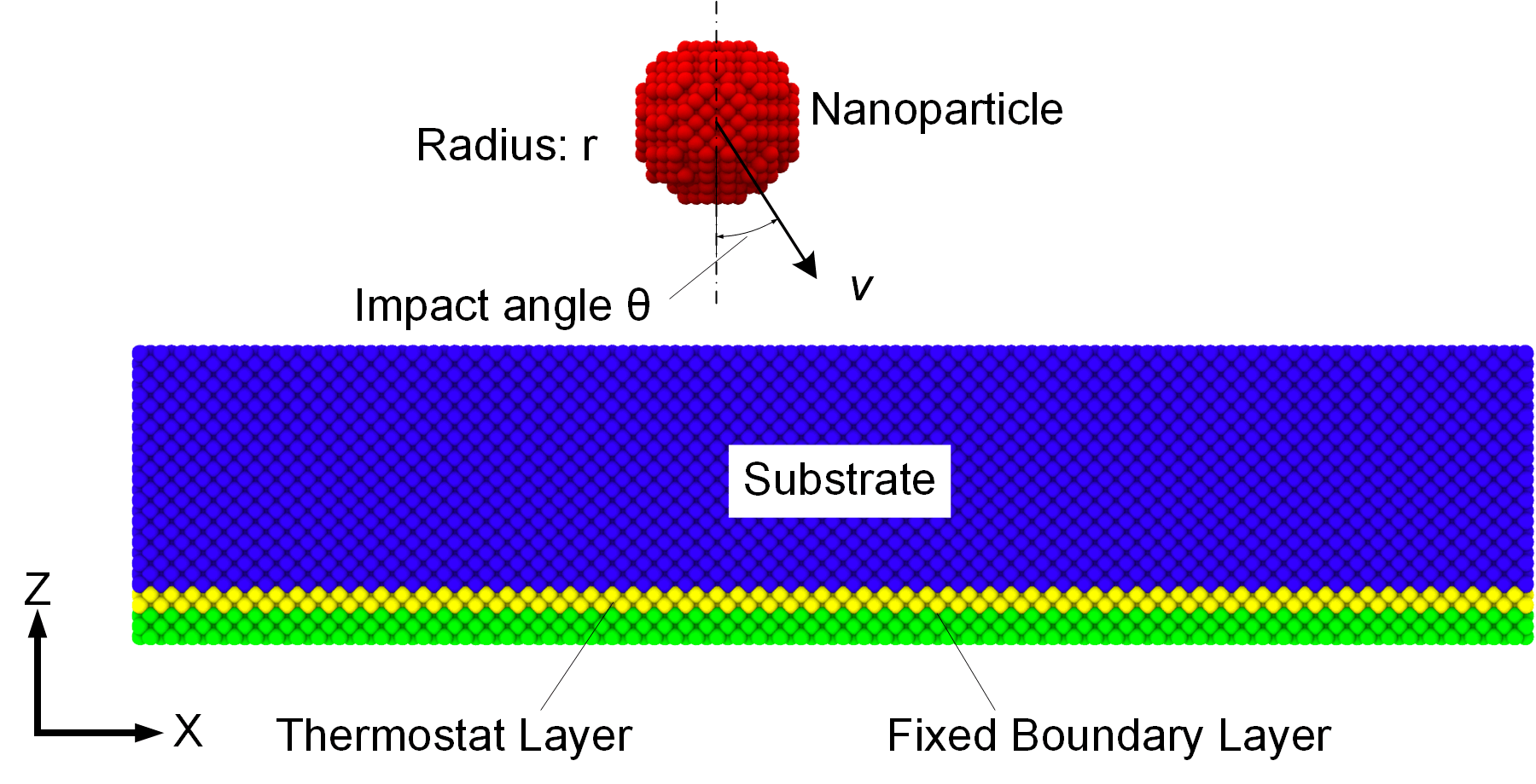}
	\caption{Schematic of MD Simulation Model of Cold spray}
	\label{fig:md-model}
\end{figure}

In this study, a classical molecular dynamics code, named Large-scale Atomic/Molecular Massively Parallel Simulator (LAMMPS), is used to simulate the cold spray process \cite{plimpton1995fast} and the atomic visualization of the MD simulation results is processed by an open source software termed Open Visualization Tool (OVITO) \cite{ovito}. This study considers the impact between a nanoparticle and a metal substrate in three-dimension (3D). The material of the simulation is copper (Cu), so both of the nanoparticle and substrate consist of Cu atoms. The substrate is made of about 250,000 atoms and the nanoparticle is made of about 1,000 atoms. Face-centered cubic (FCC) lattice structure is used. Moreover, the schematic of the MD simulation model for the cold spray process is shown in Figure\ref{fig:md-model}, where the size of the substrate is $240\AA \times 240\AA \times 50\AA$ and the radius of nanoparticle is $15 \AA$. The constant parameter of FCC lattice structure is $3.61 \AA$. More details about the parameters of MD simulation are listed in Table\ref{tab:parameters}.

\renewcommand{\arraystretch}{1.5}
\begin{table}[htbp]
	\centering
	\small
	\caption{Parameters of the cold spray MD simulation}
	\label{tab:parameters}
	\begin{tabular}{p{2cm} p{4cm} p{6cm}}
		\toprule
		\multirow{2}{2cm}{Material parameter}    & Substrate Material         & Cu $(240\AA \times 240\AA \times 50\AA)$ Approx. 240,000 atoms \\
		                                         & Nanoparticle Material      & Cu Sphere $(Radius 10\AA-20\AA)$ Approx. 500-2,000 atoms       \\ \hline
		\multirow{7}{2cm}{Simulation parameters} & Temperature                & 298$K$                                                           \\
		                                         & Potential Used             & Embedded Atom Method (EAM)                                     \\
		                                         & Initial Stand-off distance & $40\AA$                                                        \\
		                                         & Impact Velocity            & 3-12$\AA/ps$ (300-1200 $m/s$)                                                   \\
		                                         & Particle Size              & $10\AA-20\AA$                                                  \\
		                                         & Angle of Impact            & $0^{\circ}-30^{\circ}$                                           \\
		                                         & Time step                  & 0.001 ps (picoseconds)                                         \\
		                                         \bottomrule
	\end{tabular}
\end{table}

\subsection{Closed loop image aided optimization}

\begin{figure}[htbp]
	\centering
	\subfigure[before impact]{\includegraphics[width=5.5cm]{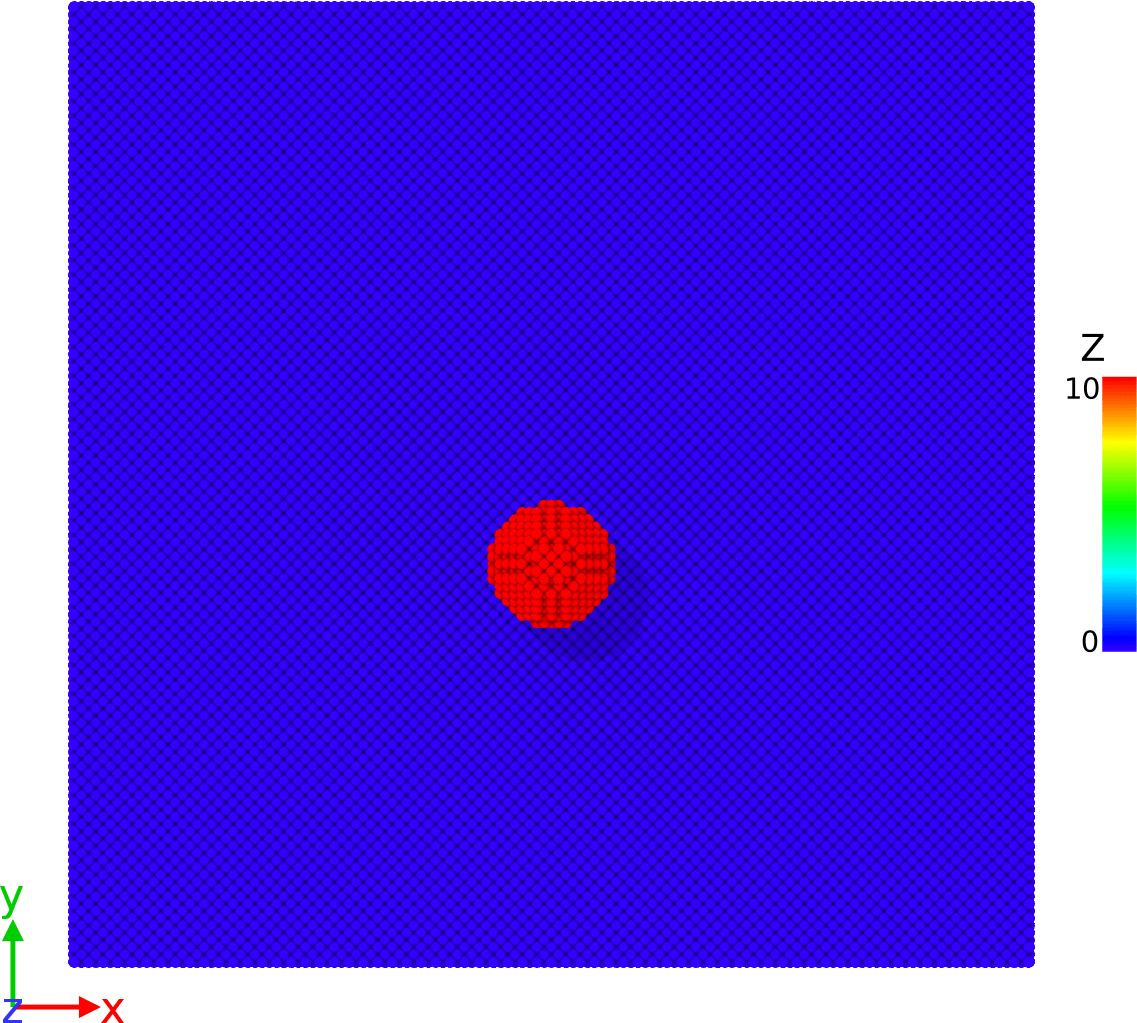}
		\label{fig:topviewleft}}
	\subfigure[after impact]{\includegraphics[width=5.5cm]{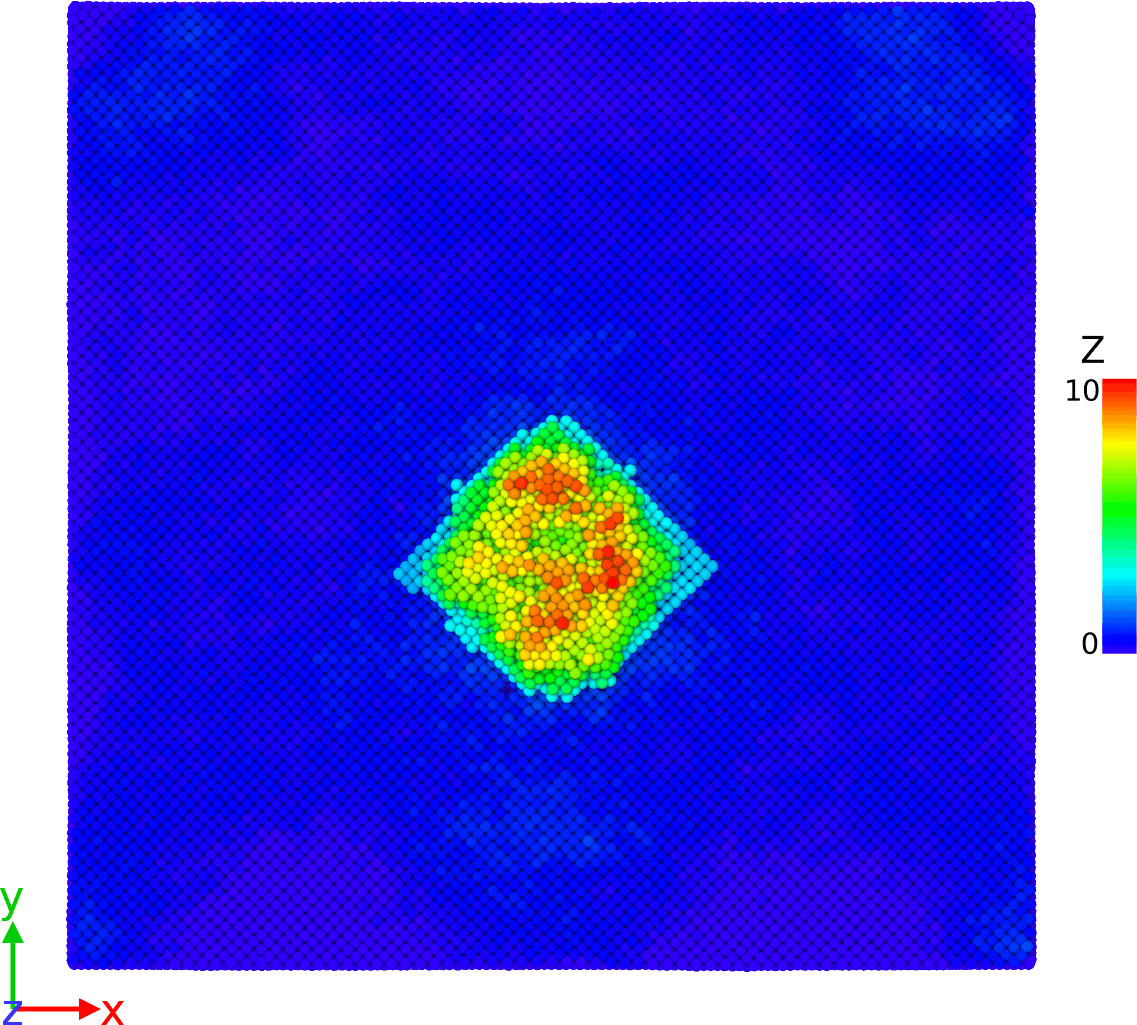}
		\label{fig:topviewright}}
	\caption{The snapshots of MD computational results on top view}
	\label{fig:topview}
\end{figure}

The CLIAO method is suggested to obtain the optimal flattening ratio $\mu$, where $\mu$ was defined as a ratio of the maximum diameter of splat (particle after impact) to original diameter of the particle before impact in the Reference\cite{jmp2018molecular}. However, it is difficult to define a diameter when the particle impact to the substrate with an angle, because the splat after impact is irregular. Therefore, a generalized flattening ratio $\mu$ is defined as the ratio of the maximum area ($S_m$) of the splat (particle after impact) to the area of the particle cross section ($S_i$) before impact in this study. Specially, the flattening ratio $\mu$ is used to measure the quality of deposition during the cold spray process. Consider an example for cold spray as shown in Figure\ref{fig:md-model}, where the radius of the particle is $15\AA$, the velocity is $8\AA/ps$, the impact angle is $0^{\circ}$. Then Figure\ref{fig:topview} shows the snapshots of the MD results before and after impact on top view, where the contour plot is according to the Z-coordinate range from $0\AA$ to $10\AA$.

\subsubsection{Optimization formulation}

As shown in Figure\ref{fig:topview}, once the snapshots of the MD computational results are obtained, then $S_i$  and $S_m$ can be obtained from Figure\ref{fig:topviewleft} and \ref{fig:topviewright} respectively. Then, the flattening ratio $\mu$ can be calculated by

\begin{equation}
	\mu=S_m/S_i.
\end{equation}

In order to improve the quality of deposition, the maximum $\mu$ should be obtained by optimization method. Thus, this optimization problem is formulated as

Minimize:
\begin{equation}
	c(v,r,\theta)=\frac{1}{\mu}=\frac{S_i}{S_m},
\end{equation}

Subject to:
\begin{gather}
	\dot{x}_i = v_i, \label{eq:11}\\
	m_i \dot{v}_i = -\triangledown_{x_i}V, \label{eq:12}\\
	S_i=f(v,r,\theta), \\
	S_m=f(v,r,\theta), \\
	300m/s \leqslant v \leqslant 1200m/s, \\
	10\AA \leqslant r \leqslant 20\AA, \\
	0^{\circ} \leqslant \theta \leqslant 30^{\circ},	
\end{gather}

\noindent where $c(v,r,\theta)$ is the objective function and $\mu$ means flattening ration mentioned above. The symbols $v, r, \theta$ denote the velocity, radius, impact angle of particles. Moreover, the cold spray process is simulated by MD, so the objective function also obeys to the Newton's law for MD as shown in equations\eqref{eq:11} and \eqref{eq:12}. It should be noted that the design variables ($v,r,\theta$) are restricted in appropriate intervals in accordance with the practical applications.

\subsubsection{Image processing technique}

As mentioned above, the image processing technique is used to obtain $S_i$ and $S_m$ from Figure\ref{fig:topview}, but how to generate the snapshots automatically is the first step need to be completed. In this study, the open source software OVITO is used as the post-processing tool. OVITO is a scientific visualization and analysis software for atomistic and particle simulation data. Specially, OVITO has a powerful Python-based scripting interface which can process and generate various snapshots 
programmatically. Therefore, a Python script is used to process and generate a series of snapshots such like Figure\ref{fig:topview}. The script can invoke program actions like a human user does in the graphical interface and it can run from the command line without any user interaction, so the snapshots can be generated automatically.

\begin{figure}[htbp]
	\centering
	\includegraphics[width=12cm]{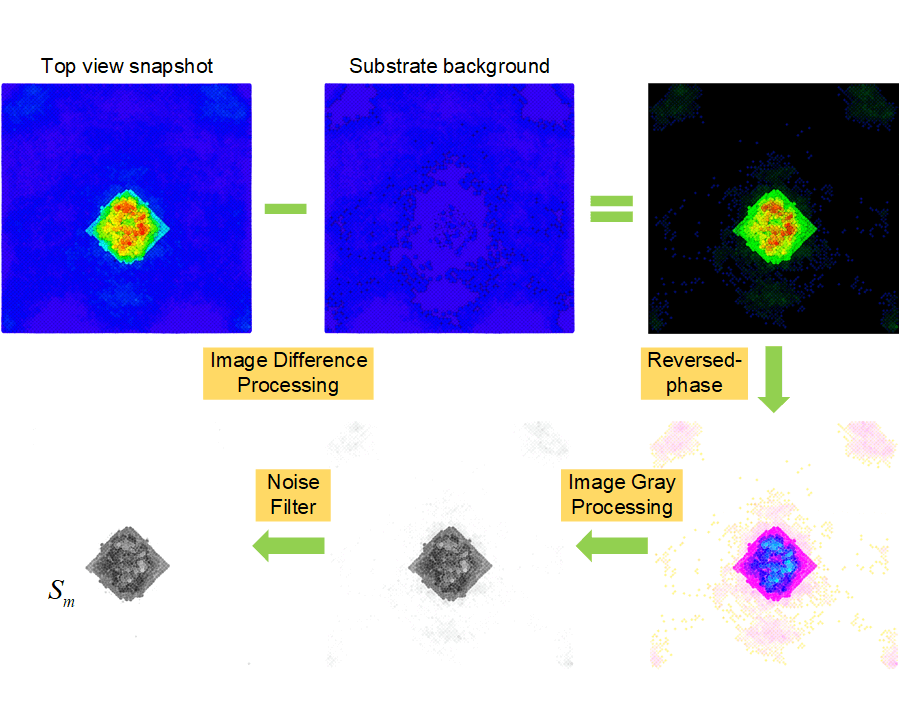}
	\caption{An example of extracting the splat from the snapshot}
	\label{fig:im-process}
\end{figure}

\begin{figure}[htbp]
	\centering
	\includegraphics[width=8cm]{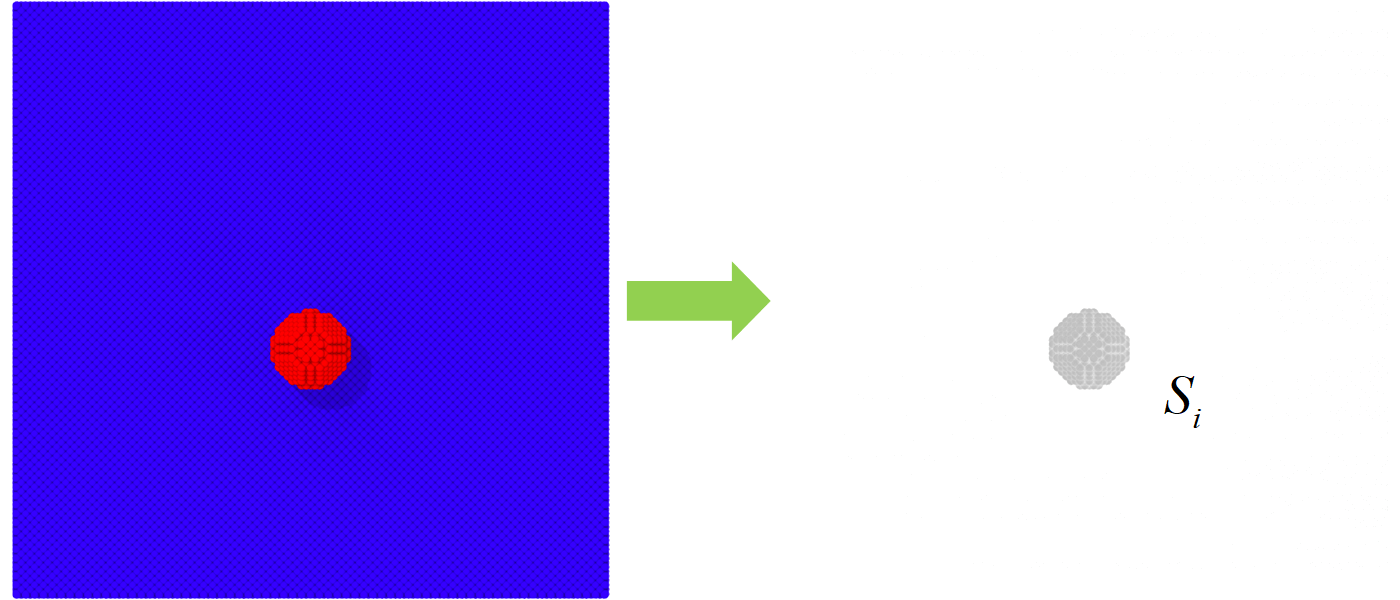}
	\caption{Separating the splat from the snapshot before impact}
	\label{fig:im-process2}
\end{figure}

Once the snapshots are generated, the next step is how to extract the splat (particle after impact) from the snapshots. Figure\ref{fig:im-process} shows the process of extracting the splat step by step. It is obvious that the top view snapshot and the substrate background image should be obtained by the Python script firstly. Then make an image difference processing to remove the substrate background from the top view snapshot, so that only the splat (usually located in the central part) will be left in the snapshot. In order to show the result more clearly in the publication, a reversed-phase operation should be applied to the result image which can convert the black background to write. It should be pointed out that this operation is not necessary for the optimization process, is just for the publishing. After that, an image gray processing should be applied to the result image so that the image can be identified clearly by computer. It is worth mentioning that there are some noises in the result image, so a noise filter is used to filter the noises in the last stage. Finally, the splat is separated from the top view snapshot as shown in Figure\ref{fig:im-process}, so that $S_m$ can be calculated by computer easily. Similarly, $S_i$ can be also obtained by the same way which as shown in Figure\ref{fig:im-process2}.

\subsubsection{BPNN assisted optimizations}

It is well known that optimization is an iterative process and the MD program need to be evoked again and again, so this is a time-consumed process. Thus, the BPNN is used to accelerate the process by building a mate-model for the objective function. The BPNN is one of the popular artificial neural networks \cite{goh1995back}. It's a type of multi-layered feed-forward neural network which can handle nonlinear problem flexibly \cite{zhang1998forecasting}. The BPNN has been widely used in many fields and algorithms, such as genetic algorithm (GA) \cite{irani2011evolving}, PSO \cite{zhang2007hybrid} and so on.

\begin{figure}[hbp]
	\centering
	\includegraphics[width=10cm]{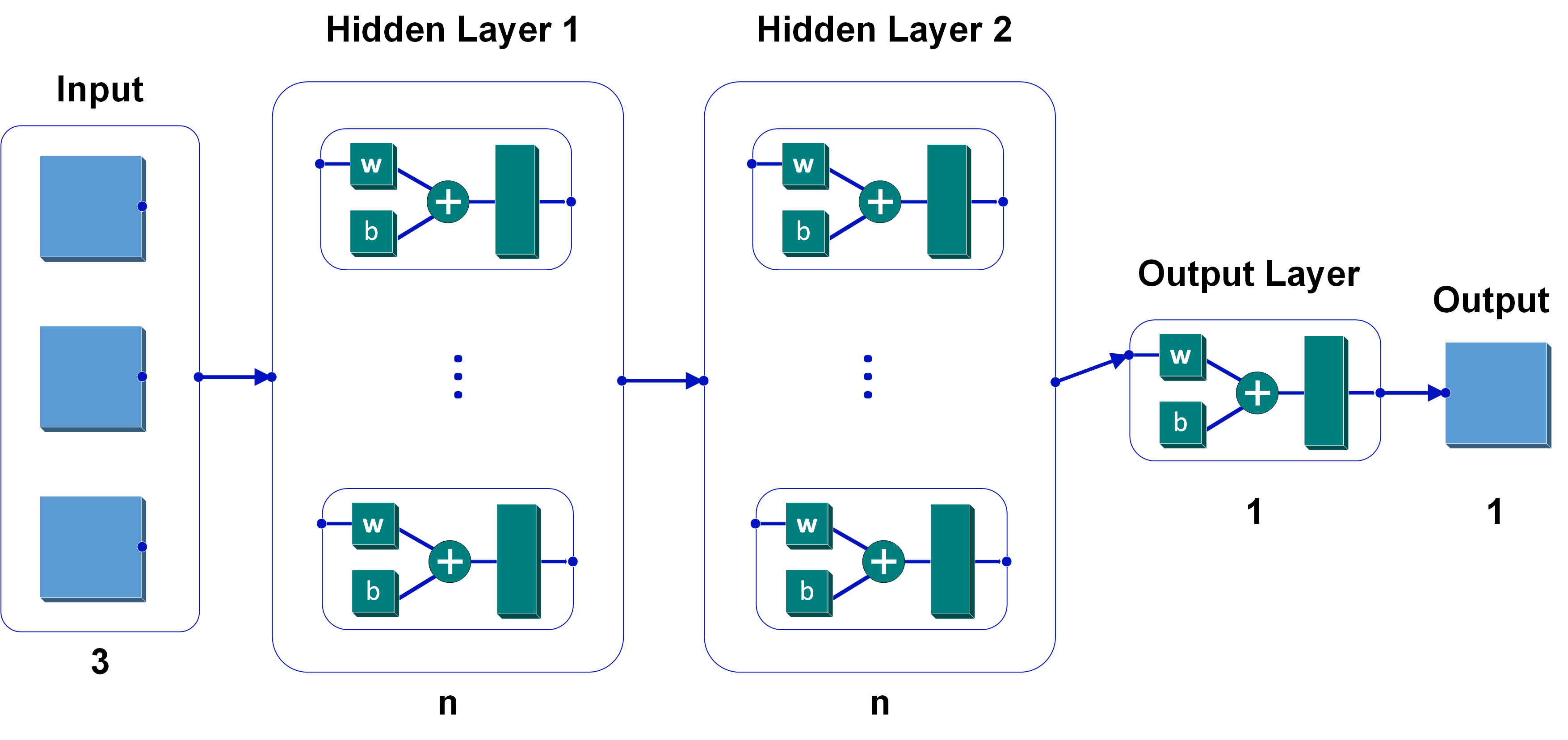}
	\caption{The construction of a BPNN}
	\label{fig:bpnn}
\end{figure}

In this study, the BPNN is used to forecast the value of the objective function, so that the optimal solution can be found more efficiently than the classic optimization loop. Generally, a BPNN is made up of an input layer, an output layer and several hidden layers as shown in Figure\ref{fig:bpnn} and the details can be found in the Referance\cite{zhang2002evaluation}. Furthermore, in order to forecast the value of the objective function, a set of sample data should be used to construct a BPNN, then the BPNN need to be trained,and finally the trained BPNN can be used to forecast the value of objective function.

\section{Results and discussions}

Several popular optimization algorithms including KRG-EGO, PSO, DE are tested and some comparisons are made in this section, where the EGO is started with 20 sample points while PSO and DE is started with 20 particles. All the program is executed under the Linux system and the MD simulation is calculated by parallel LAMMPS with 24 MPI processor. The image processing is realized by Python script and the optimization algorithms are running in Matlab R2017a.

\subsection{Results of classic optimization methods}

\begin{figure}[htbp]
	\centering
	\includegraphics[width=8cm]{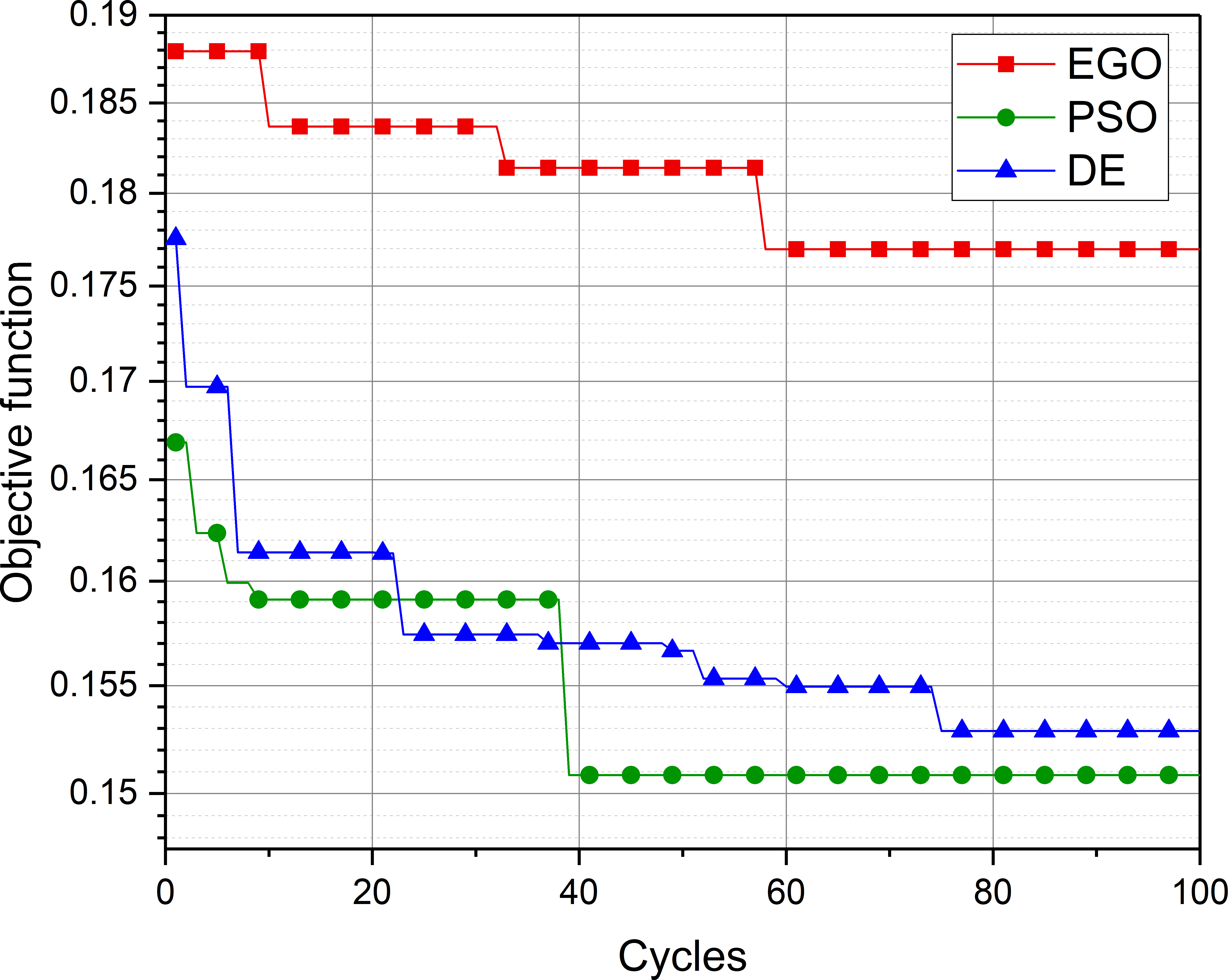}
	\caption{Convergence curve of the objective function in the optimization procedure by different methods}
	\label{fig:classicopt}
\end{figure}

\begin{figure}[htbp]
	\centering
	\subfigure[EGO]{\includegraphics[width=3.9cm]{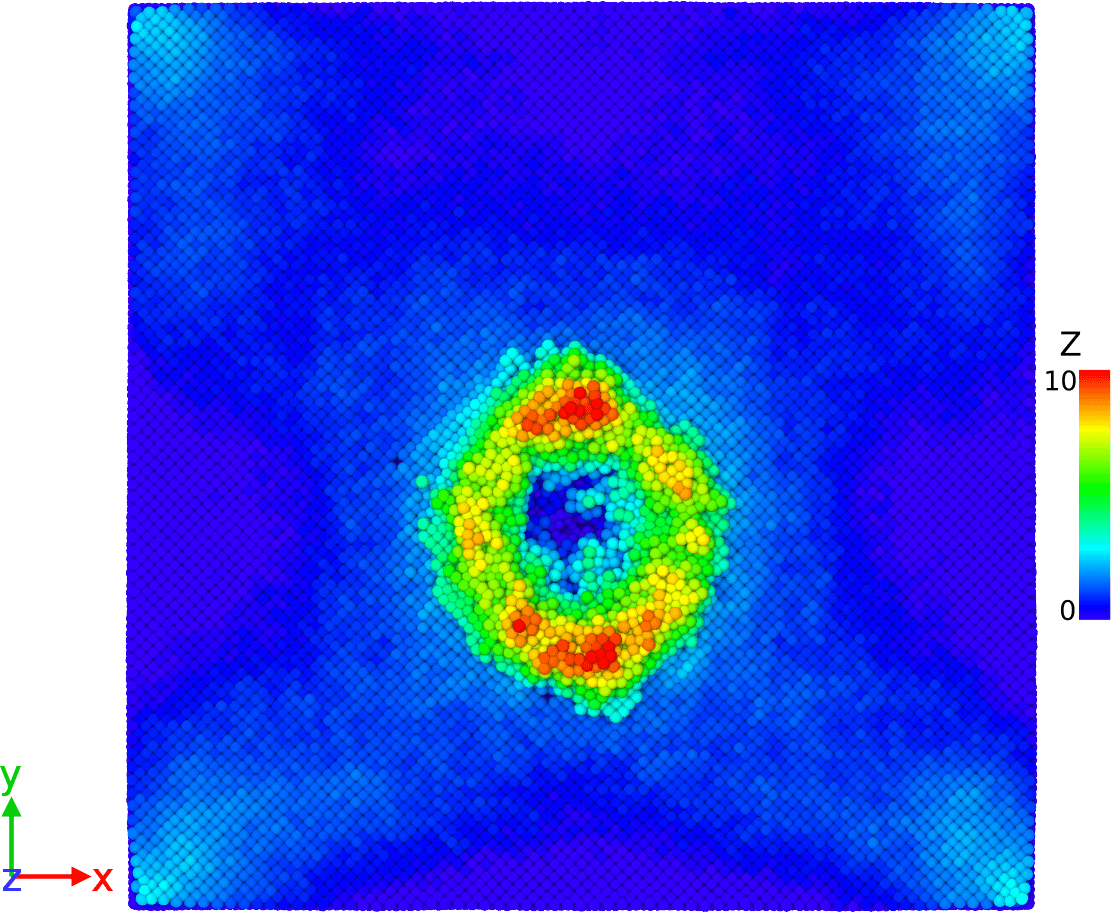}}
	\subfigure[PSO]{\includegraphics[width=3.9cm]{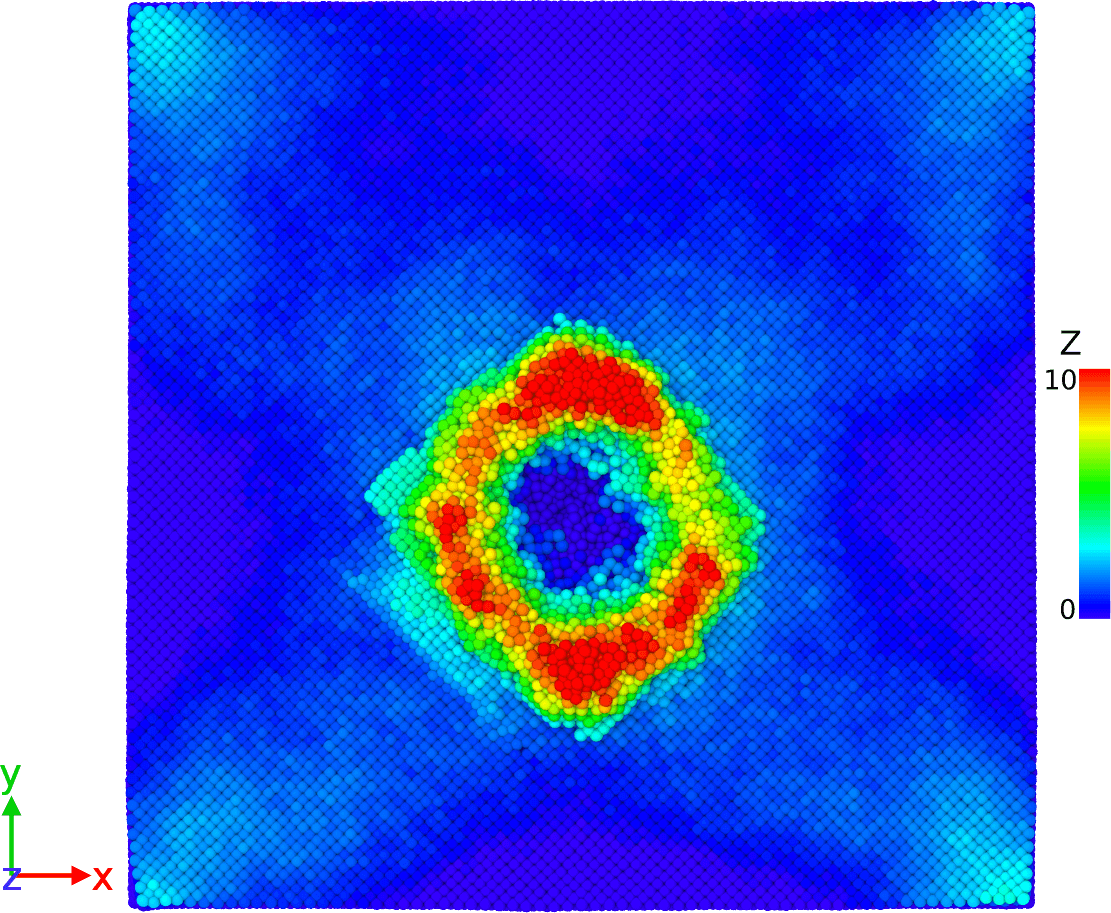}}
	\subfigure[DE]{\includegraphics[width=3.9cm]{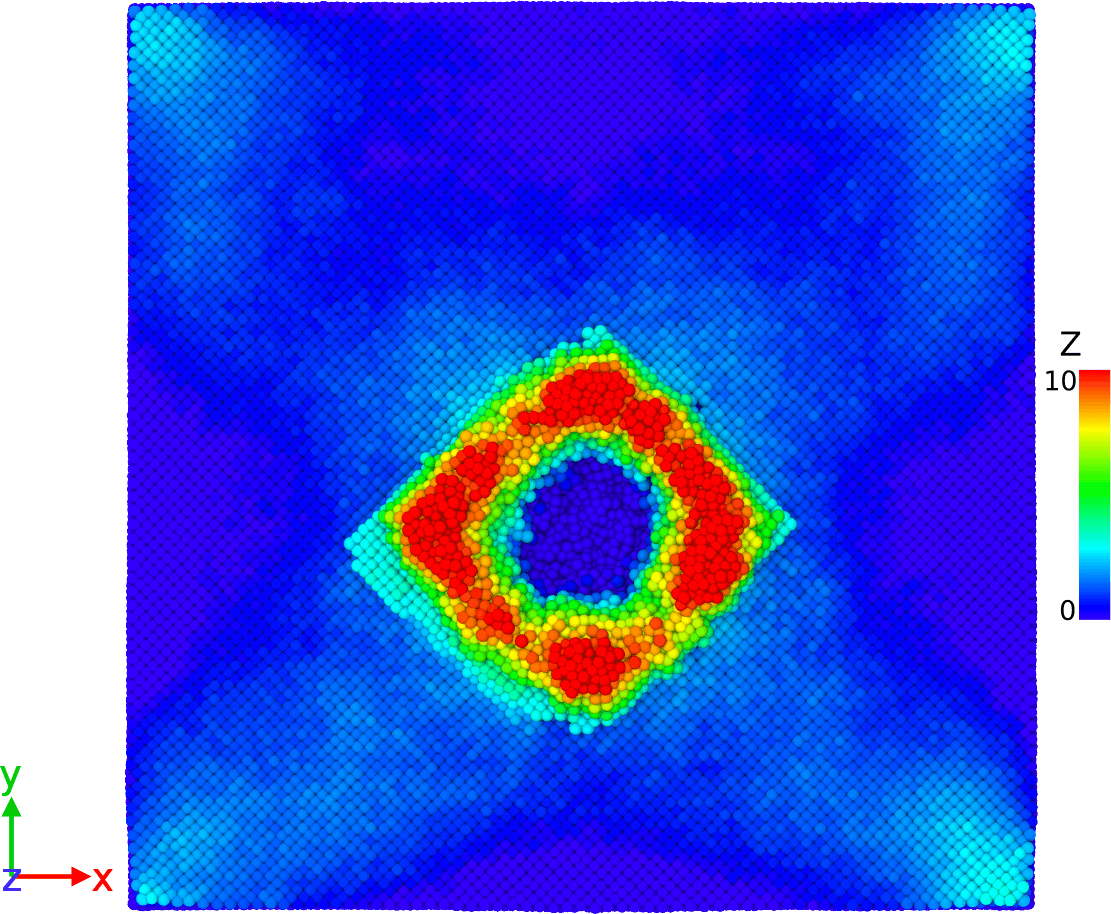}}
	\caption{The comparison of the MD simulation results after impact between classic optimizations}
	\label{fig:topcla}
\end{figure}

\begin{figure}[htbp]
	\centering
	\subfigure[$0ps$]{\includegraphics[width=3.9cm]{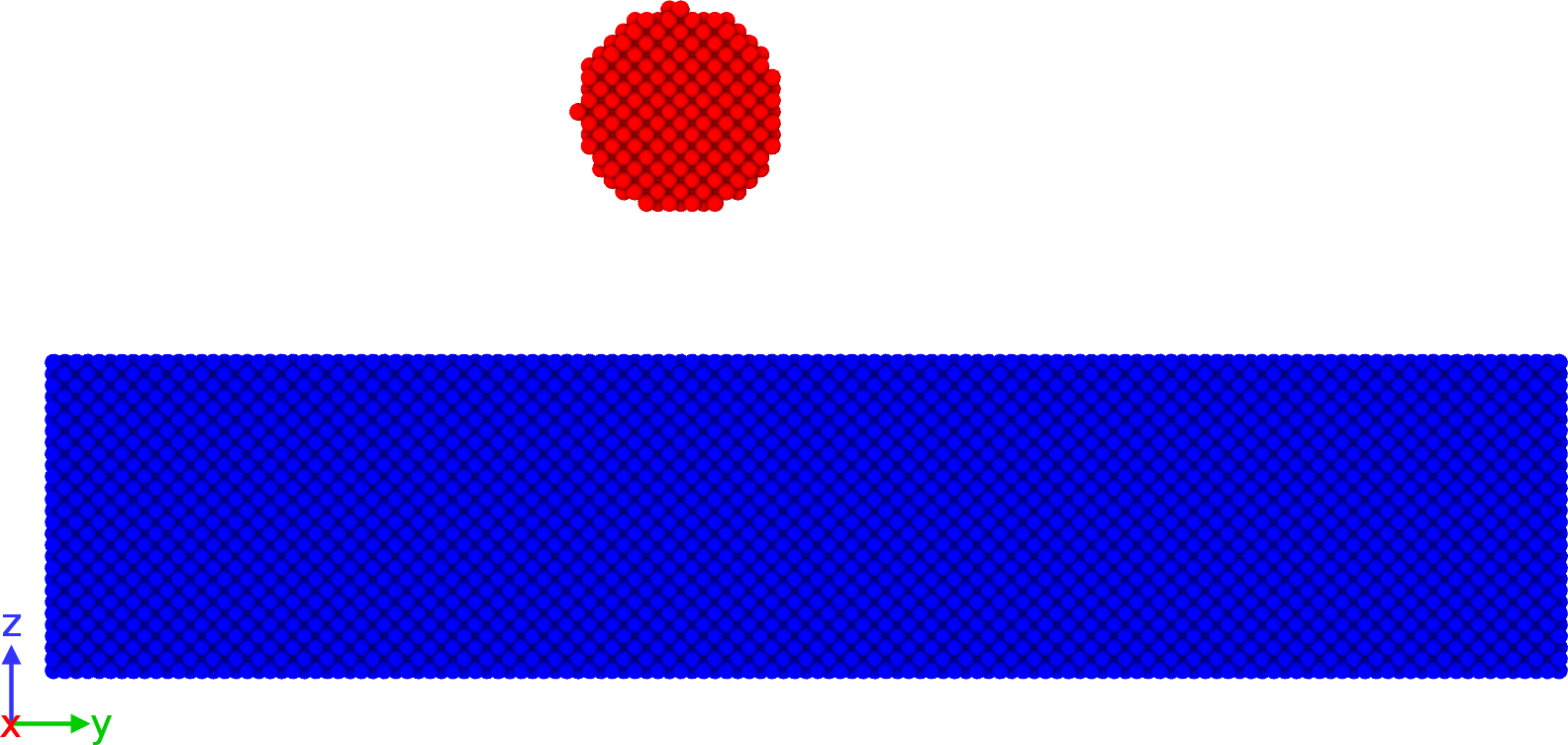}}
	\subfigure[$1ps$]{\includegraphics[width=3.9cm]{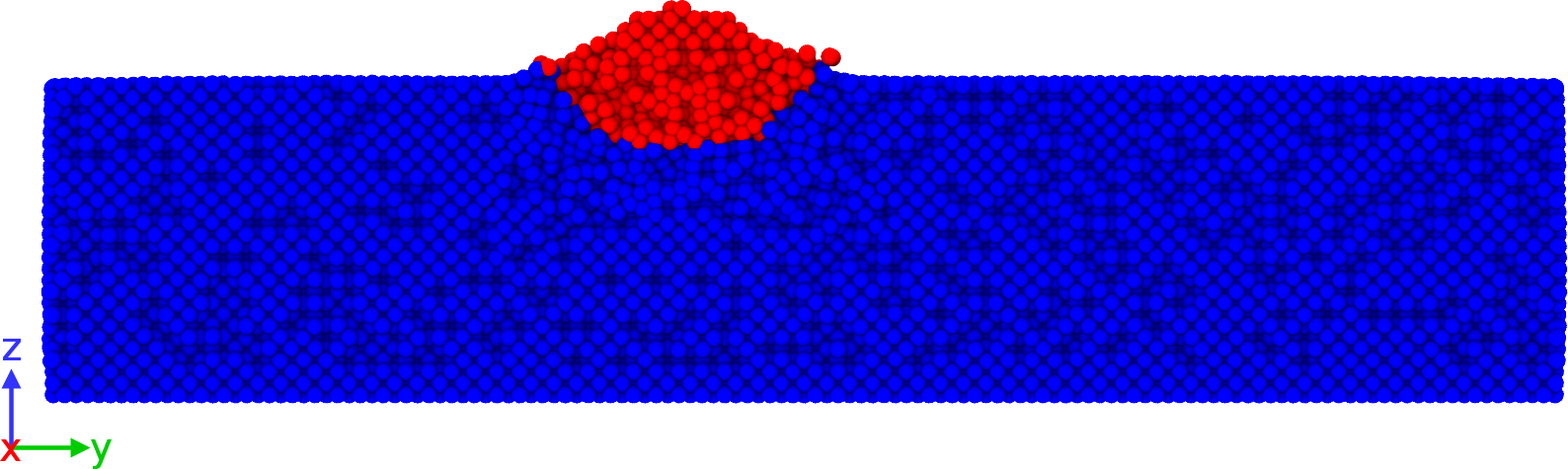}}
	\subfigure[$1.5ps$]{\includegraphics[width=3.9cm]{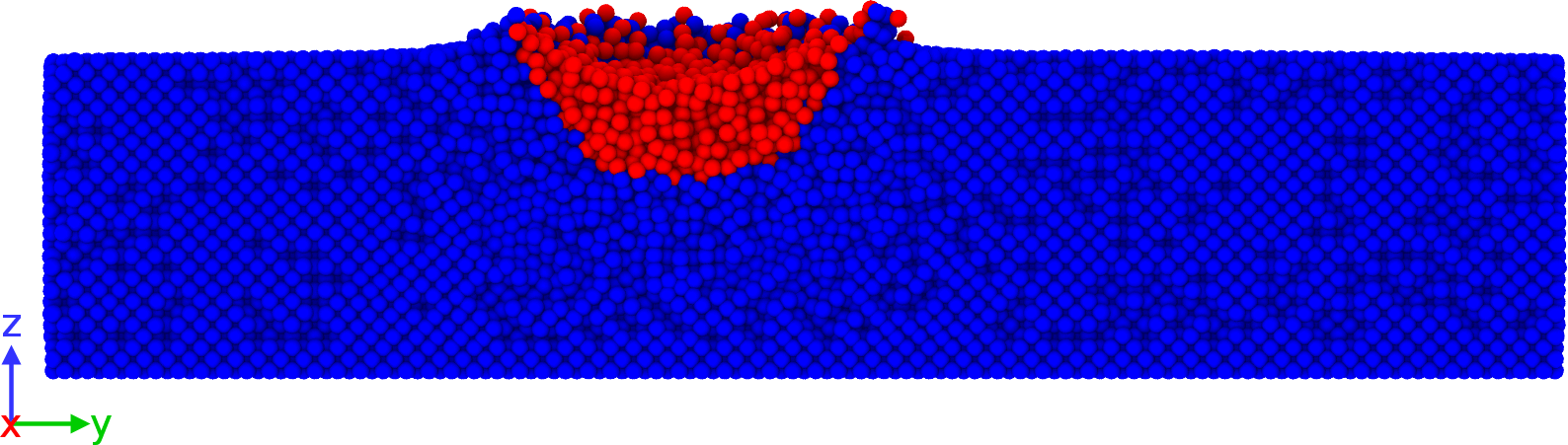}}
	
	\subfigure[$3ps$]{\includegraphics[width=3.9cm]{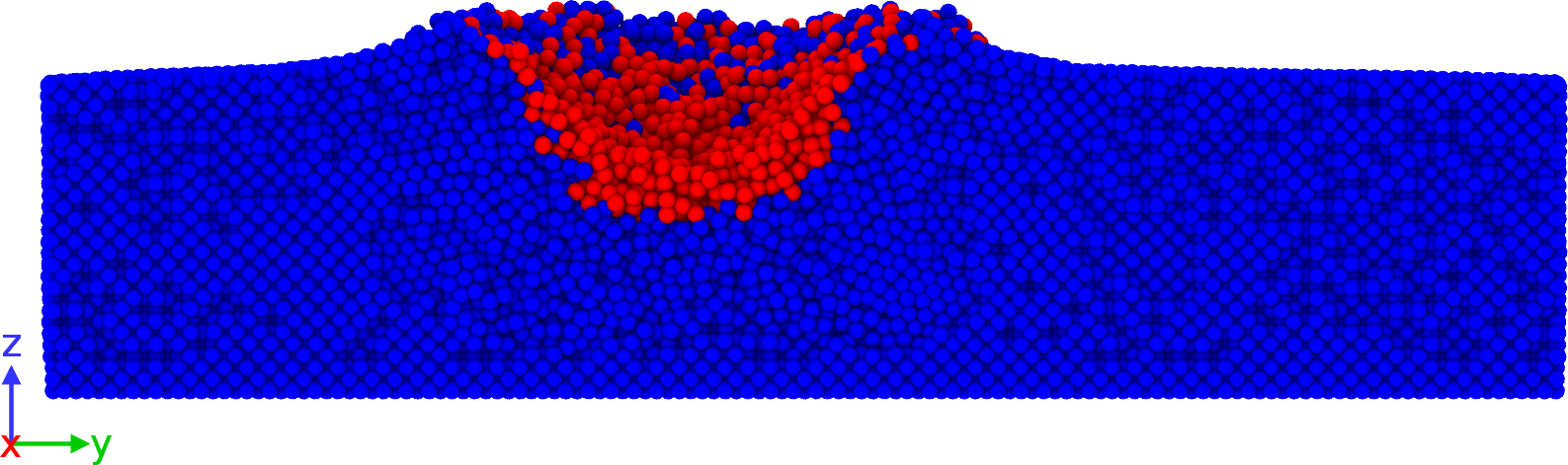}}
	\subfigure[$6ps$]{\includegraphics[width=3.9cm]{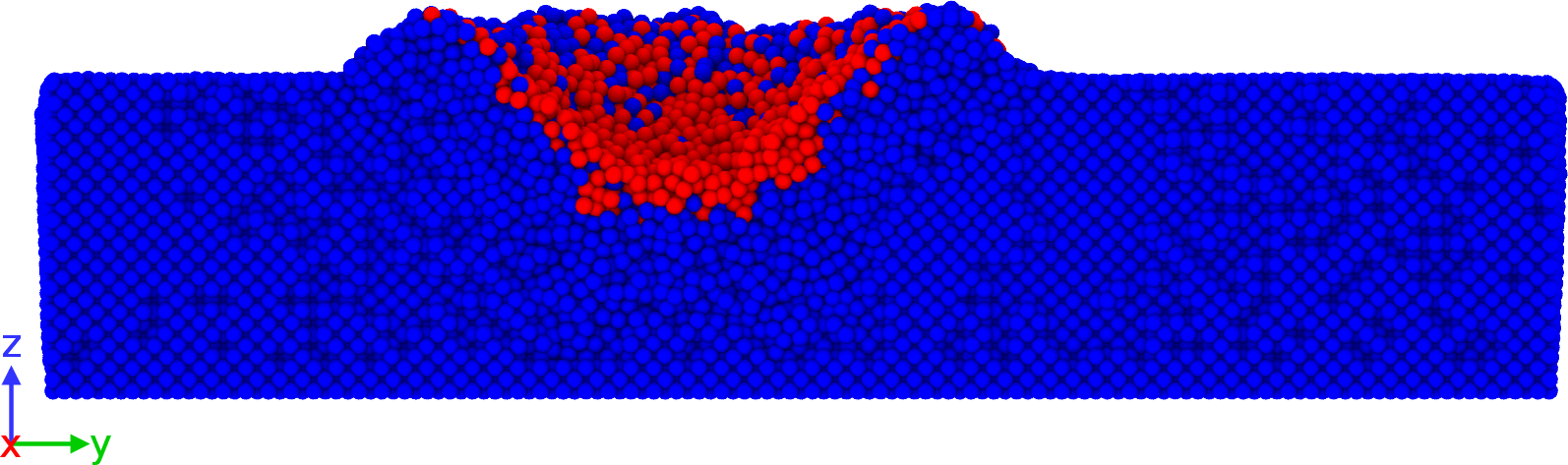}}
	\subfigure[$9ps$]{\includegraphics[width=3.9cm]{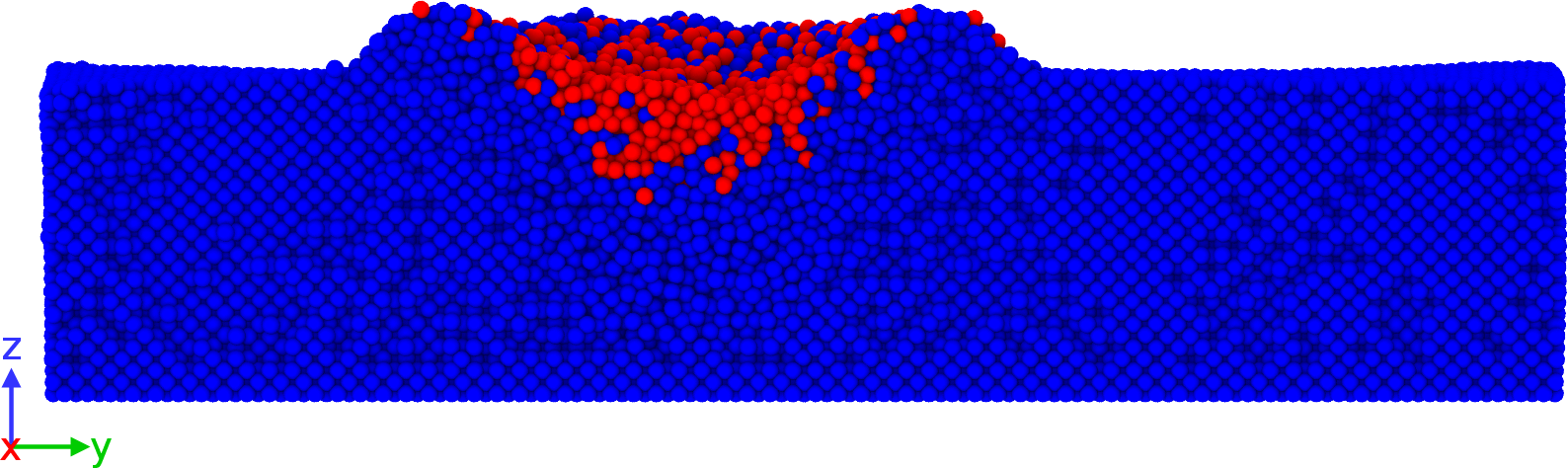}}
	\caption{The MD simulation snapshots for the optimal solution (PSO)}
	\label{fig:mds}
\end{figure}

\begin{figure}[htbp]
	\centering
	\subfigure[$1ps$]{\includegraphics[width=3.9cm]{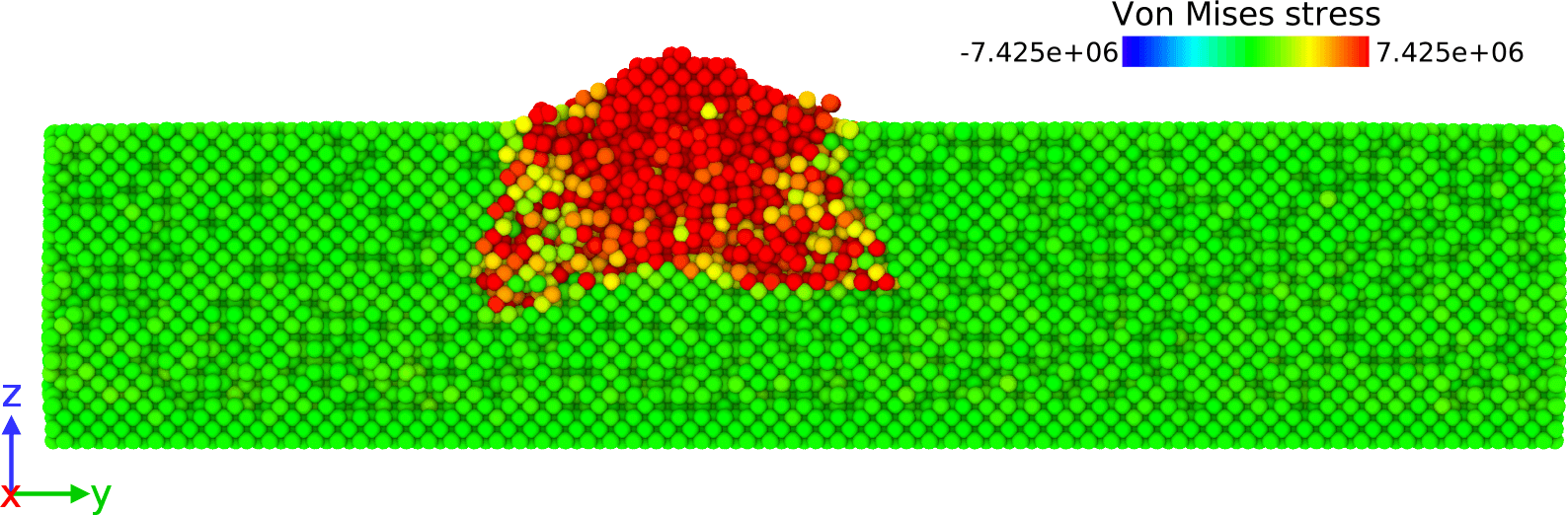}}
	\subfigure[$1.5ps$]{\includegraphics[width=3.9cm]{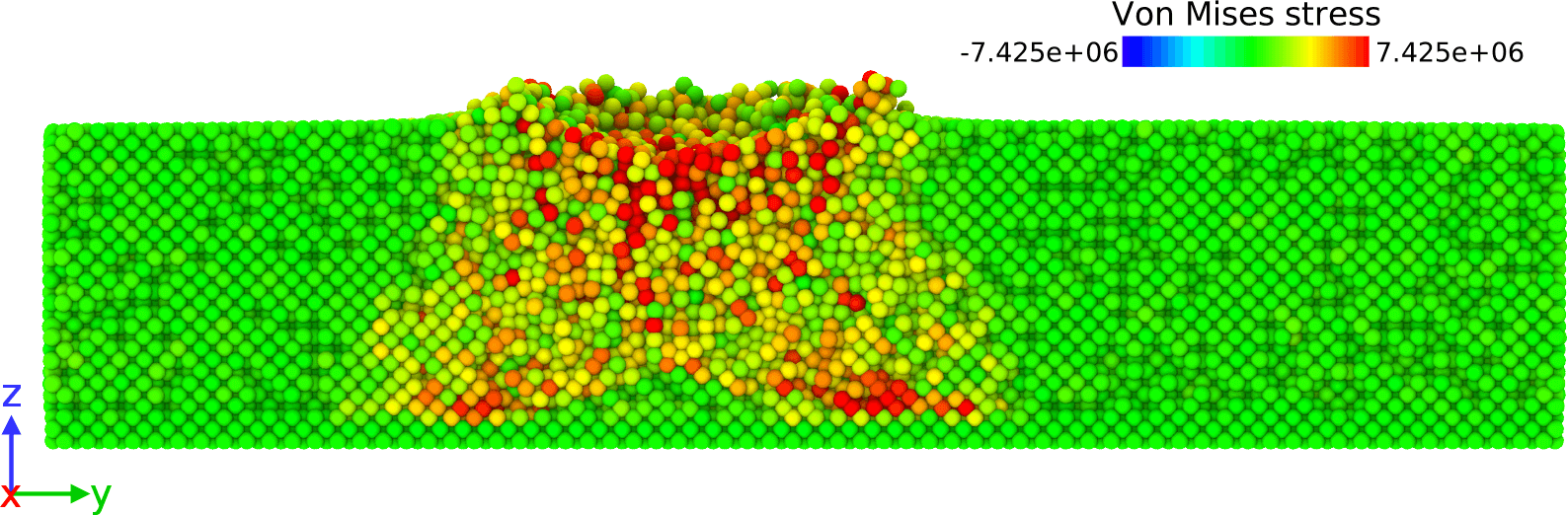}}
	\subfigure[$6ps$]{\includegraphics[width=3.9cm]{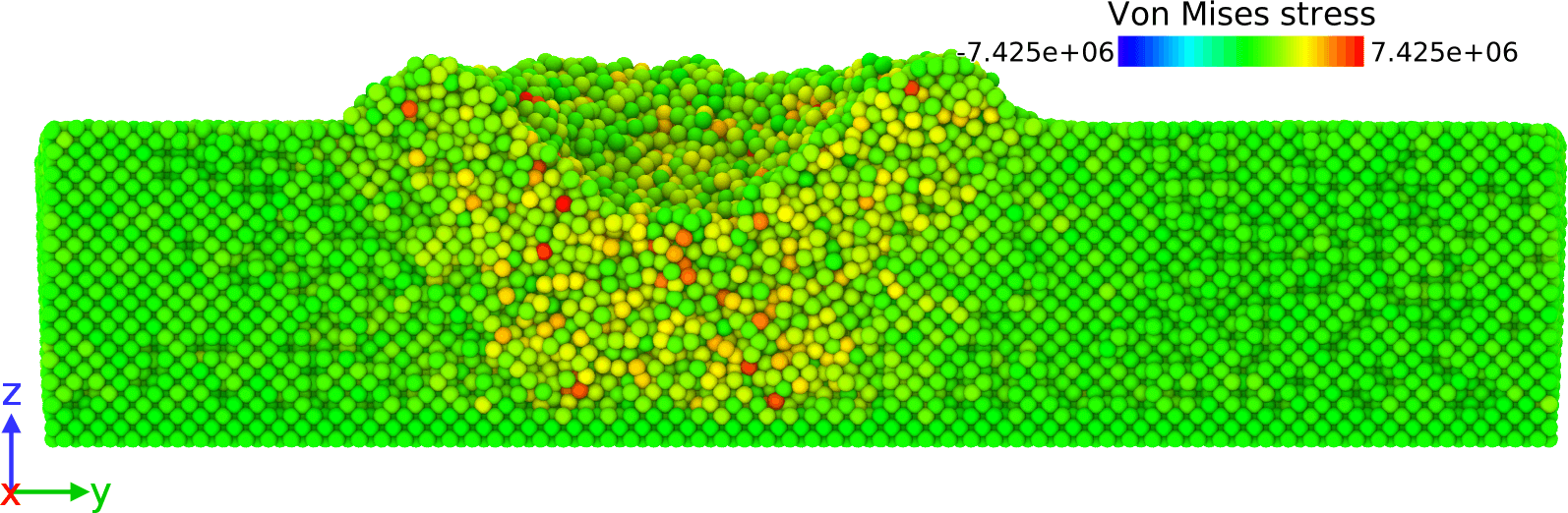}}
	\caption{The Von Mises stress contour plots for the optimal solution (PSO)}
	\label{fig:cst}
\end{figure}

As shown in Figure\ref{fig:framework}, both the classic and the BPNN assisted optimization methods are included and compared in this study. The convergence curves of the objective function in the optimization procedure are shown in Figure\ref{fig:classicopt}. It can be found that each of these methods can obtain an available solution and behaves great convergent tendency. Specially, EGO is a surrogate optimization method while other two are heuristic algorithms, and it begins with a Kriging surrogate which based on the initial sample set. Then it adds new sample points to the sample set to improve the present optimum and usually the new sample points will be generated one by one according to Kriging surrogate in the single-point EGO algorithm. Thus, the computational cost of EGO is much lower than other two methods. Moreover, it is obvious that the PSO method obtained a better solution than other two methods and converged faster than DE. In a word, the heuristic algorithms (PSO, DE) can obtain a better solution than surrogate optimization methods (EGO) in this problem, but the EGO method is much more efficient than other two methods. The optimal solution of the above three methods is listed in Table\ref{tab:classicsol}, where $1\AA/ps=100m/s$. The comparison of the MD simulation results after impact between above three methods is shown in Figure\ref{fig:topcla}, where the contour plot is according to the Z-coordinate range from $0\AA$ to $10\AA$. Furthermore, the MD simulation snapshots for the optimal solution which obtained by PSO method are shown in Figure\ref{fig:mds} and the Von Mises stress contour plots are shown in Figure\ref{fig:cst}.

\begin{table}[htbp]
	\centering
	\small
	\caption{Optimal solution of the classic optimization methods}
	\label{tab:classicsol}
	\begin{tabular}{c c c c c}
		\toprule
		\multirow{2}{*}{Optimization} &    \multicolumn{3}{c}{Design variables}     & \multirow{2}{*}{Objective function} \\ \cline{2-4}
		                              & $v(\AA/ps)$ & $r(\AA)$ & $\theta(^{\circ})$ &                                     \\ \midrule
		             EGO              & 12.000      & 14.736   & 0.000              & 0.17697                             \\
		             PSO              & 12.000      & 16.145   & 0.706              & 0.15084                             \\
		             DE               & 11.854      & 16.219   & 0.000              & 0.15288                             \\ \bottomrule
	\end{tabular}
\end{table}

\subsection{Results of BPNN assisted optimization methods}

\begin{figure}[htbp]
	\centering
	\includegraphics[width=8cm]{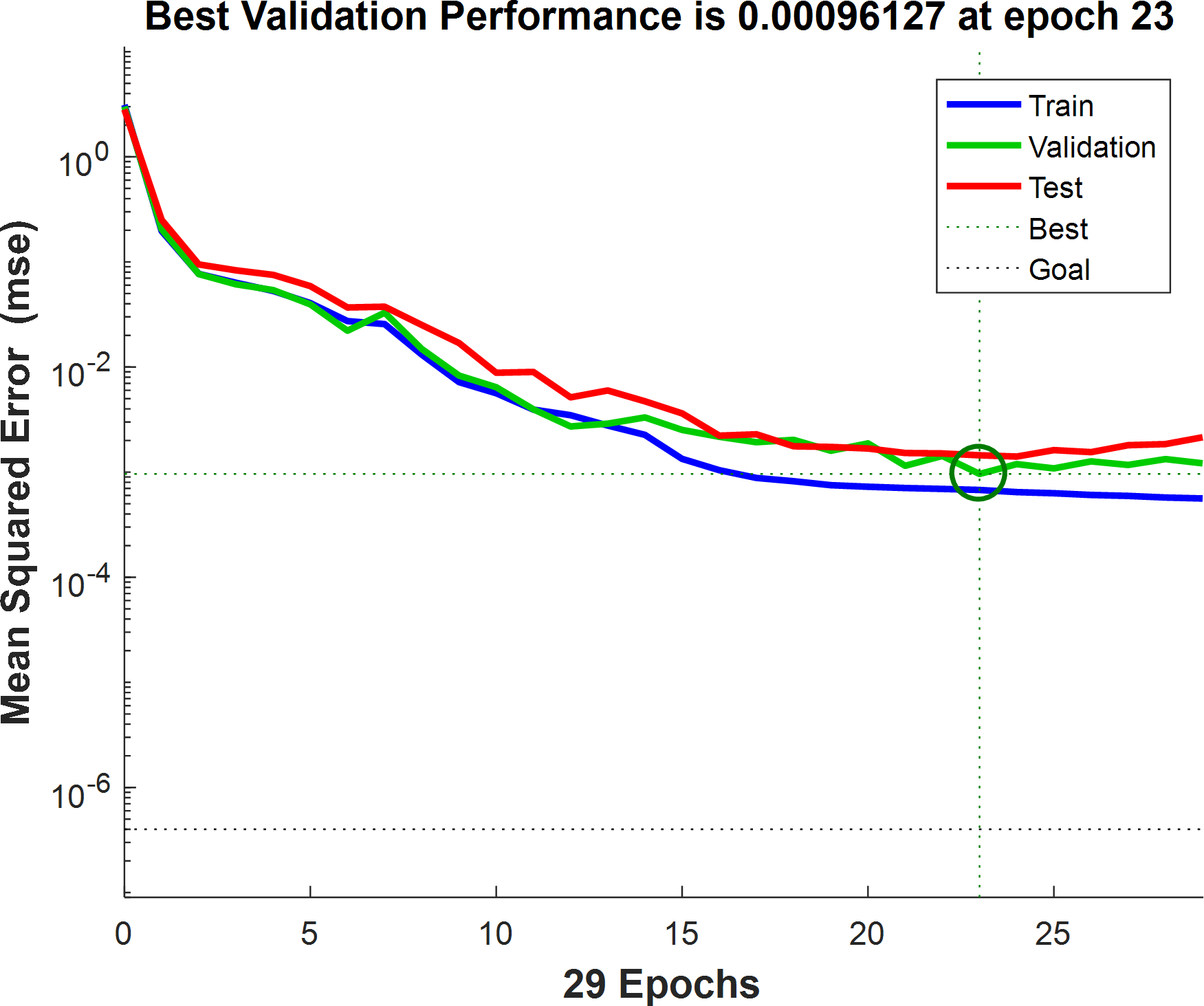}
	\caption{Validation performance of the BPNN model}
	\label{fig:bpnnvp}
\end{figure}

\begin{figure}[htbp]
	\centering
	\includegraphics[width=8cm]{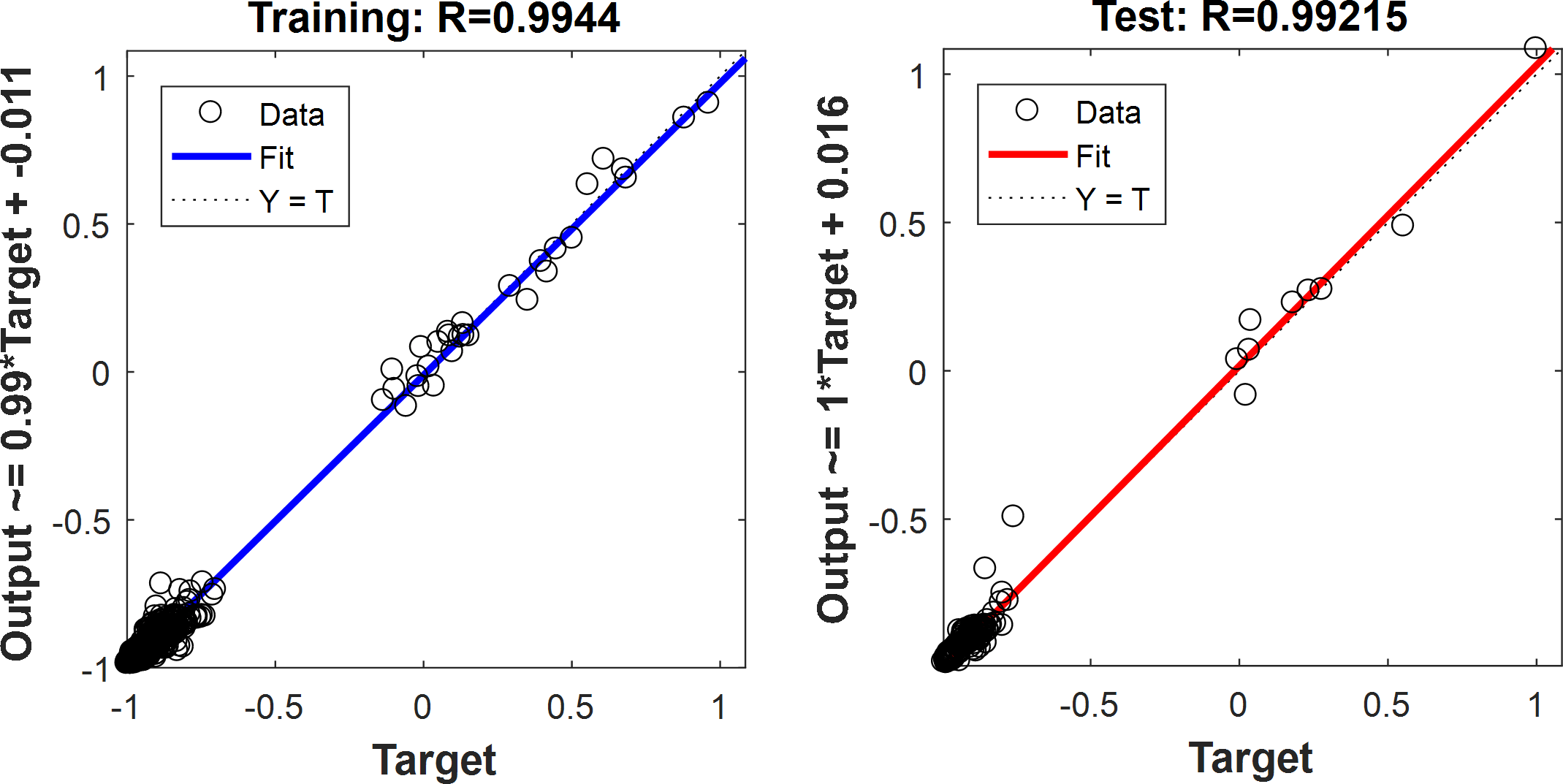}
	\caption{Regression of the BPNN model}
	\label{fig:bpnnreg}
\end{figure}

As mentioned above, the BPNN is used to accelerate the process of optimization by building a mate-model for the objective function in this study. Thus, 1000 samples are obtained firstly and then used to construct the BPNN model by training, where a triple hidden layer BPNN is used and each hidden layer has 5 nodes. After that, another 1000 samples should be used to test the performance of the BPNN model. The validation performance plot is shown in Figure \ref{fig:bpnnvp}. The regression of the trained BPNN model is shown in Figure \ref{fig:bpnnreg}, and the left is the regression of training samples while the right is the regression of testing samples. Obviously, the performance of the BPNN model is acceptable. 

\begin{figure}[htbp]
	\centering
	\includegraphics[width=8cm]{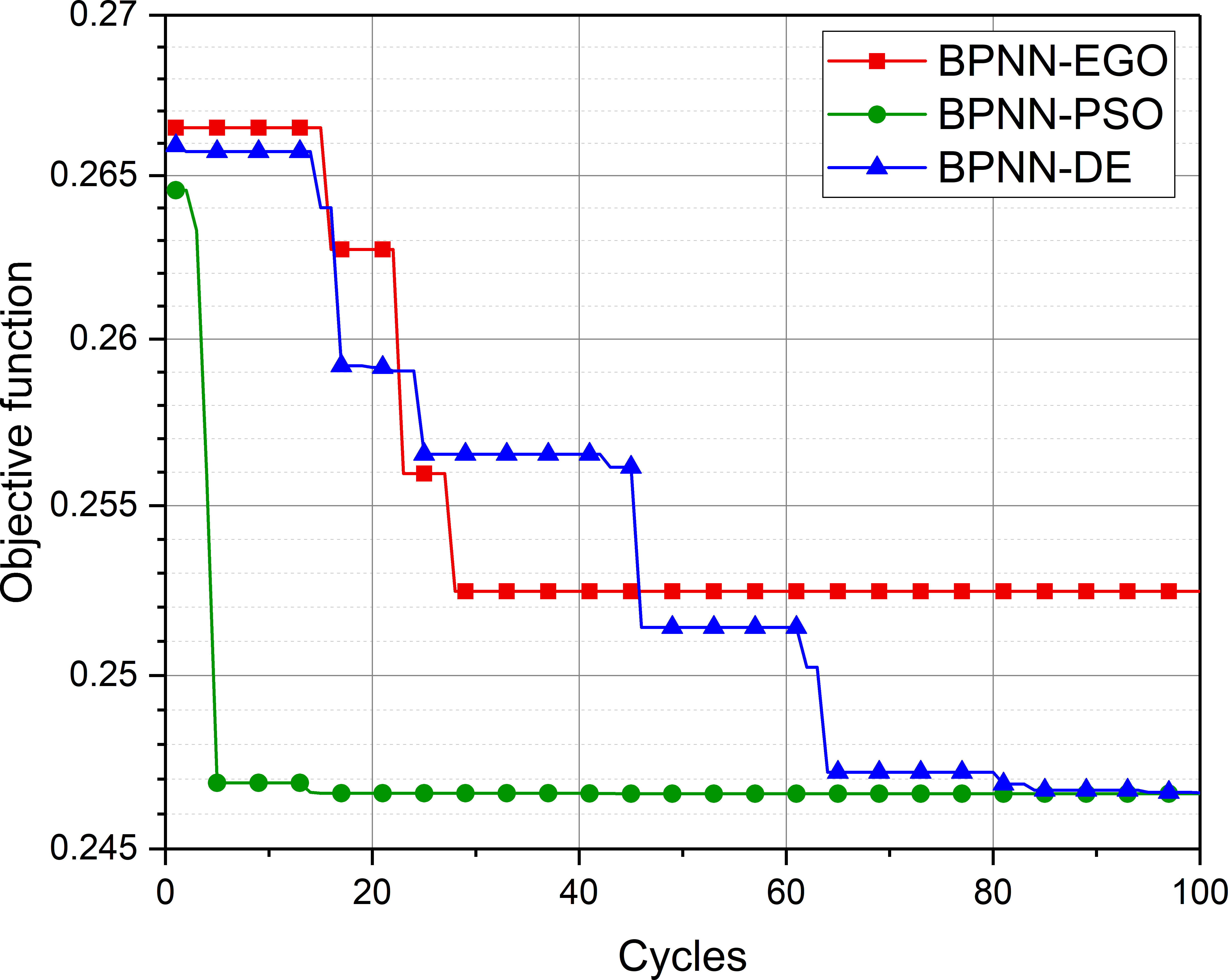}
	\caption{Convergence curve of the objective function in the bpnn assisted optimization procedures}
	\label{fig:bpnnopt}
\end{figure}

\begin{figure}[htbp]
	\centering
	\subfigure[BPNN-EGO]{\includegraphics[width=3.9cm]{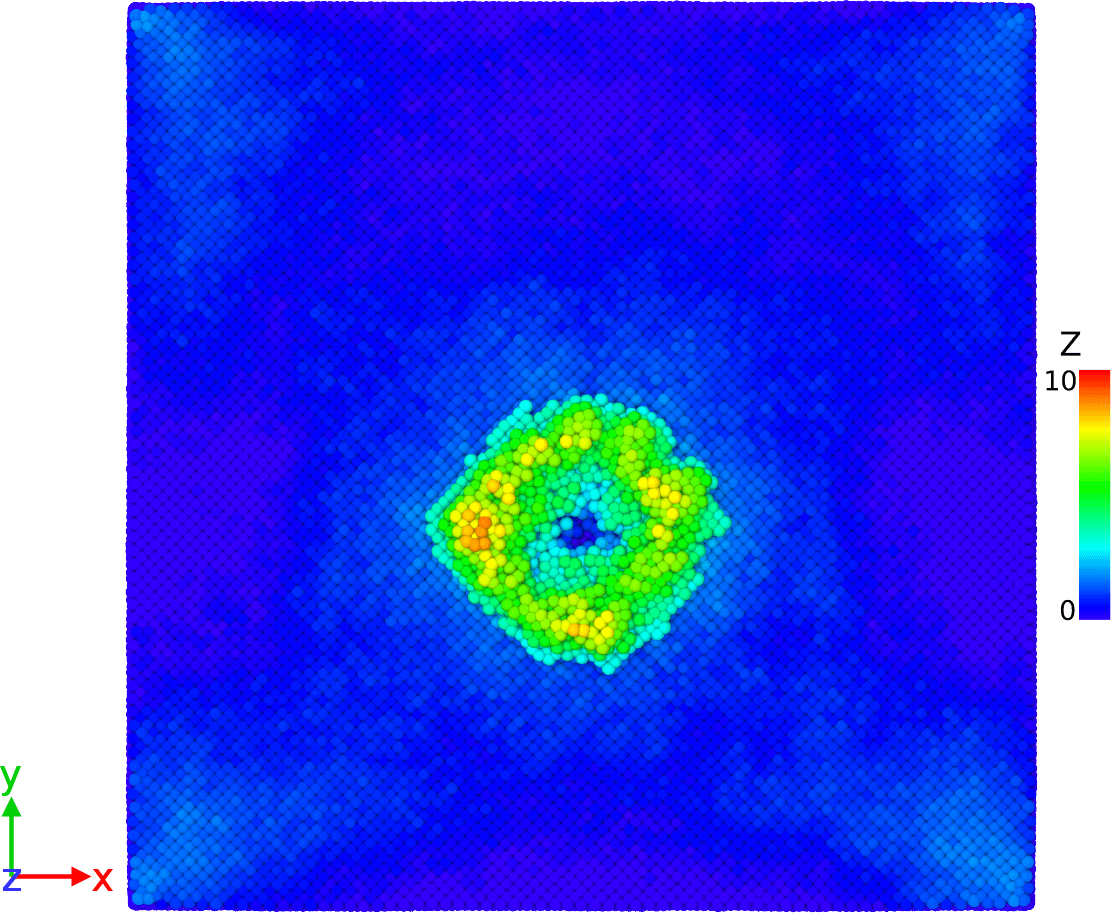}}
	\subfigure[BPNN-PSO]{\includegraphics[width=3.9cm]{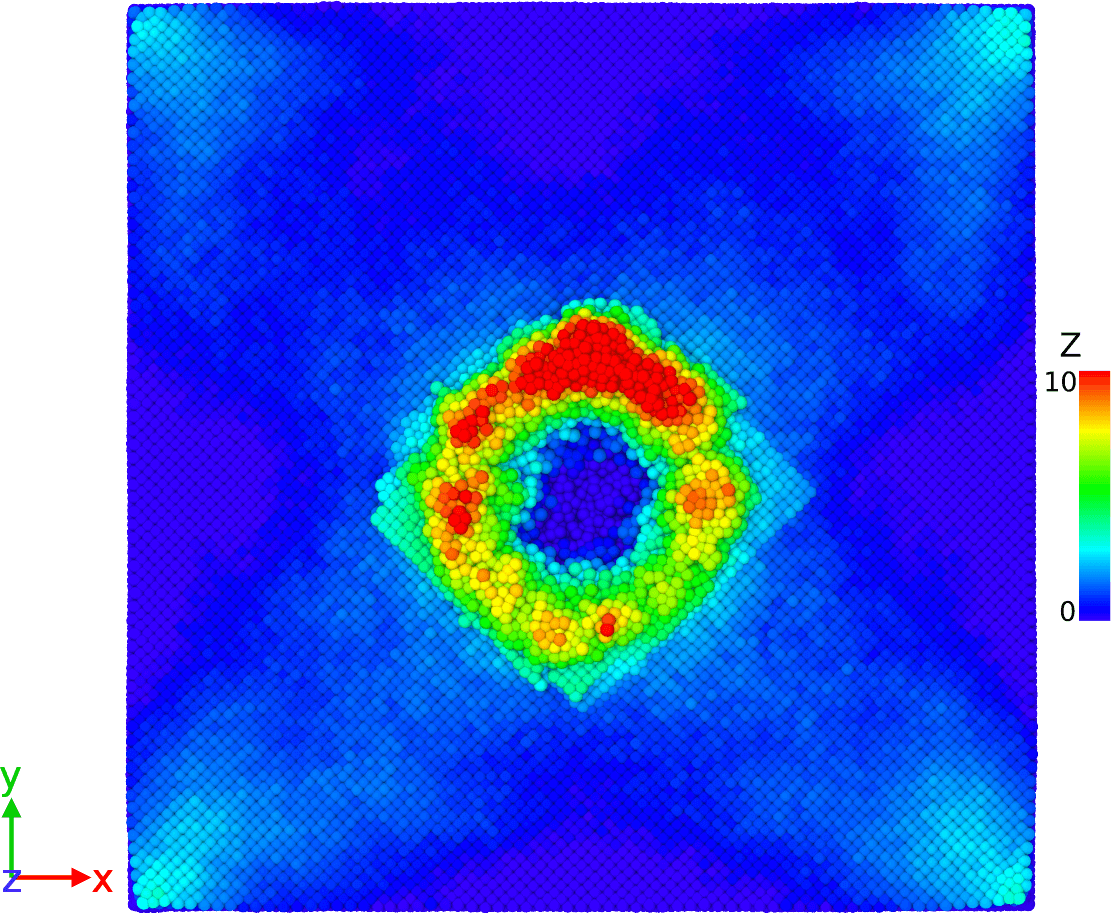}}
	\subfigure[BPNN-DE]{\includegraphics[width=3.9cm]{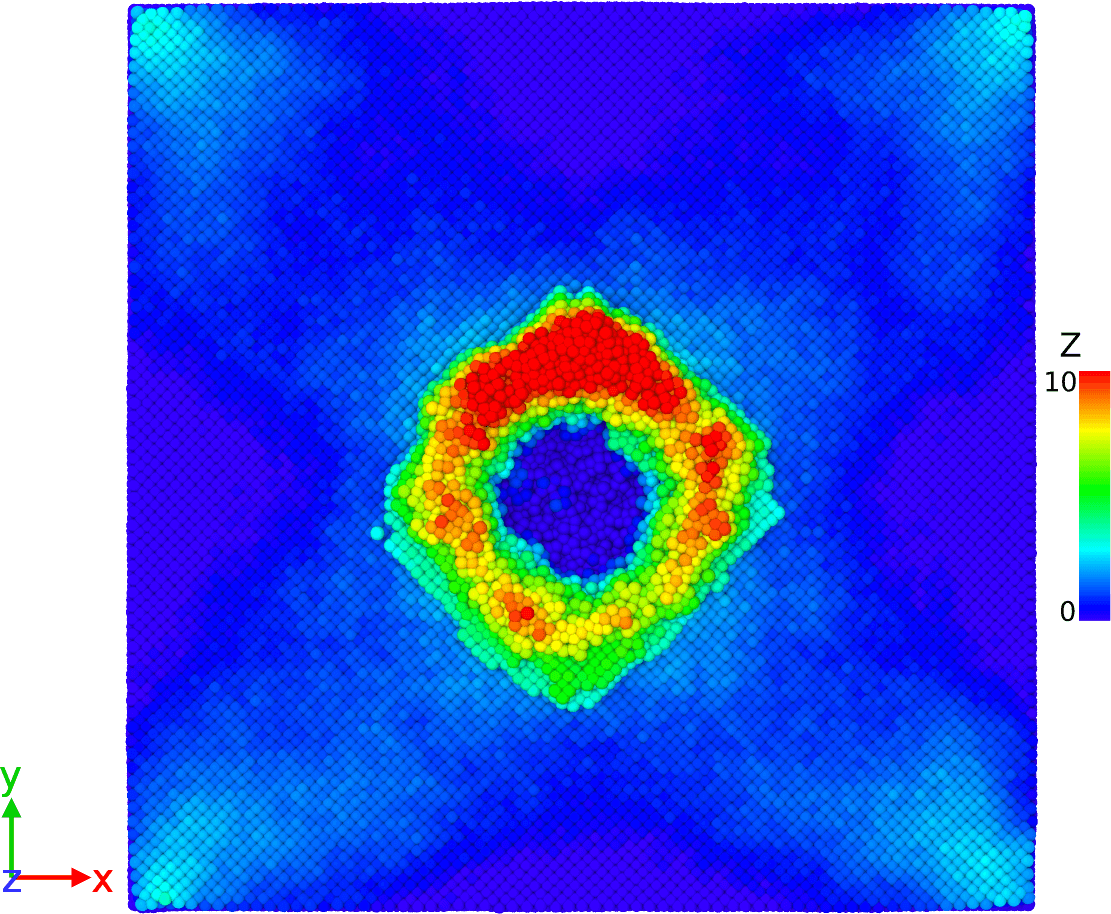}}
	\caption{The comparison of the MD simulation results after impact between bpnn assisted optimizations}
	\label{fig:topbpnn}
\end{figure}

\begin{figure}[htbp]
	\centering
	\subfigure[$0ps$]{\includegraphics[width=3.9cm]{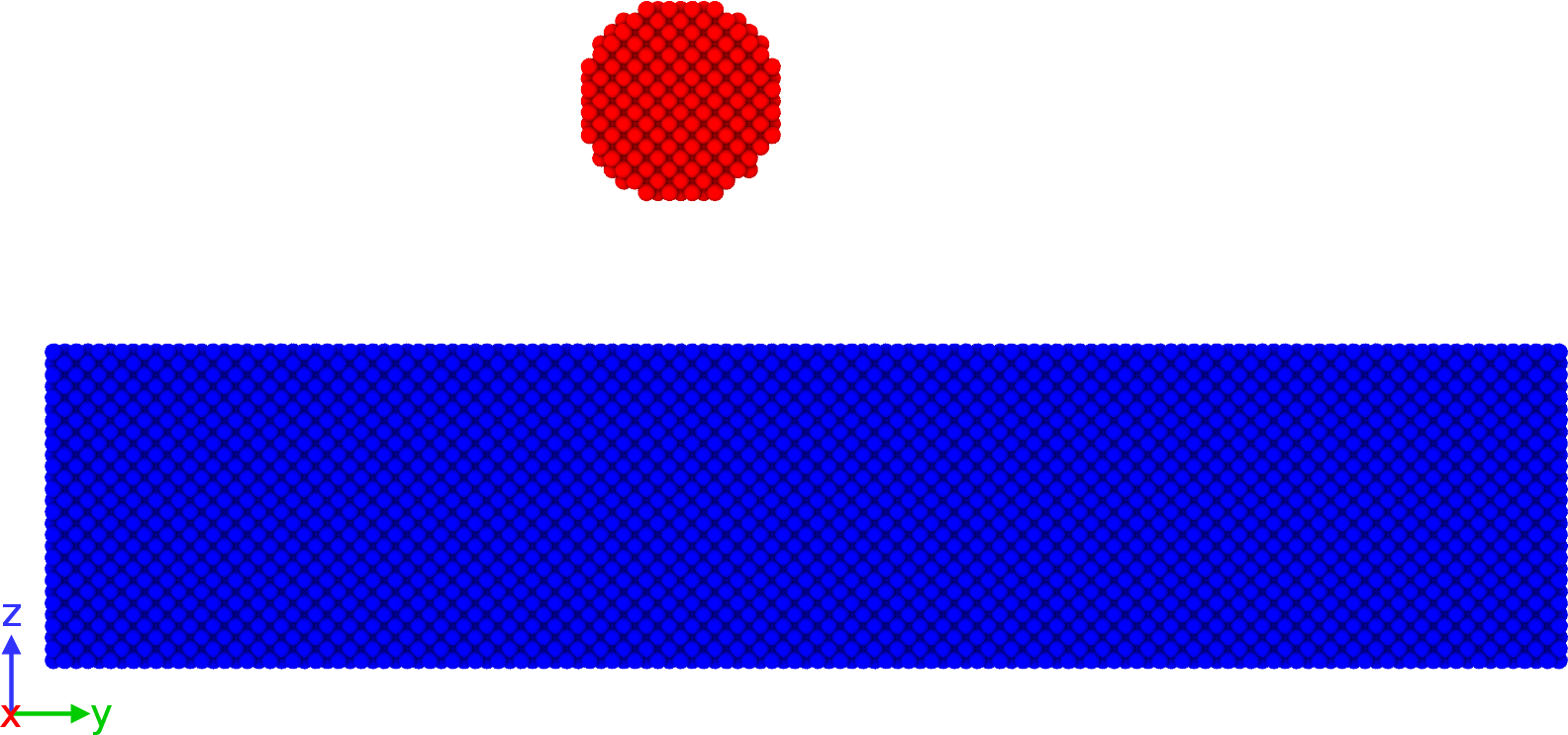}}
	\subfigure[$1ps$]{\includegraphics[width=3.9cm]{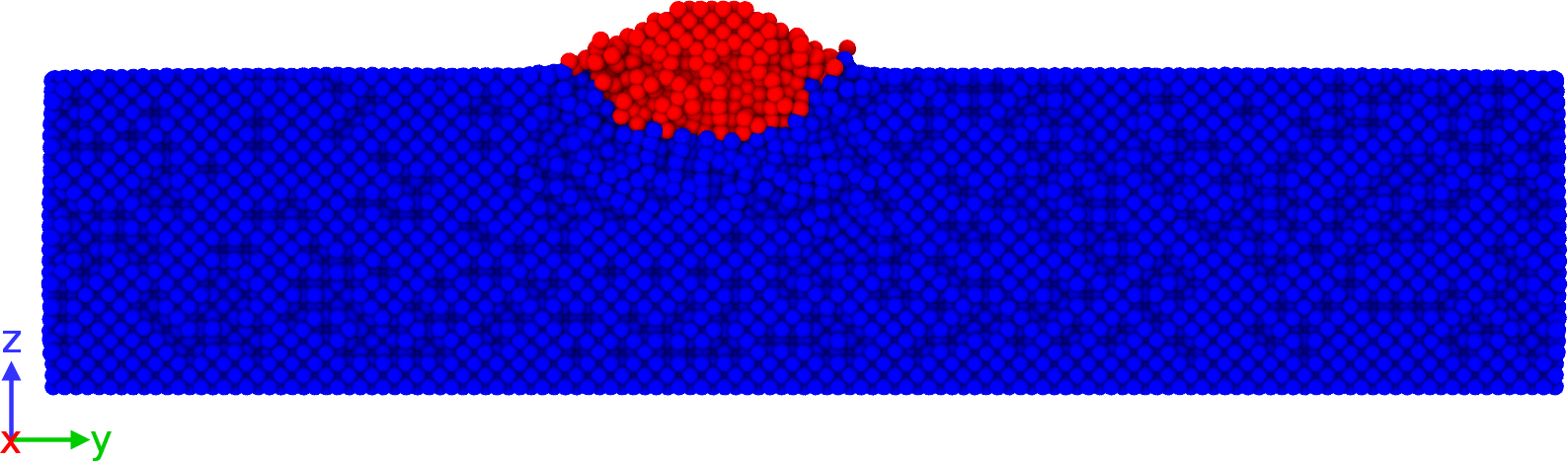}}
	\subfigure[$1.5ps$]{\includegraphics[width=3.9cm]{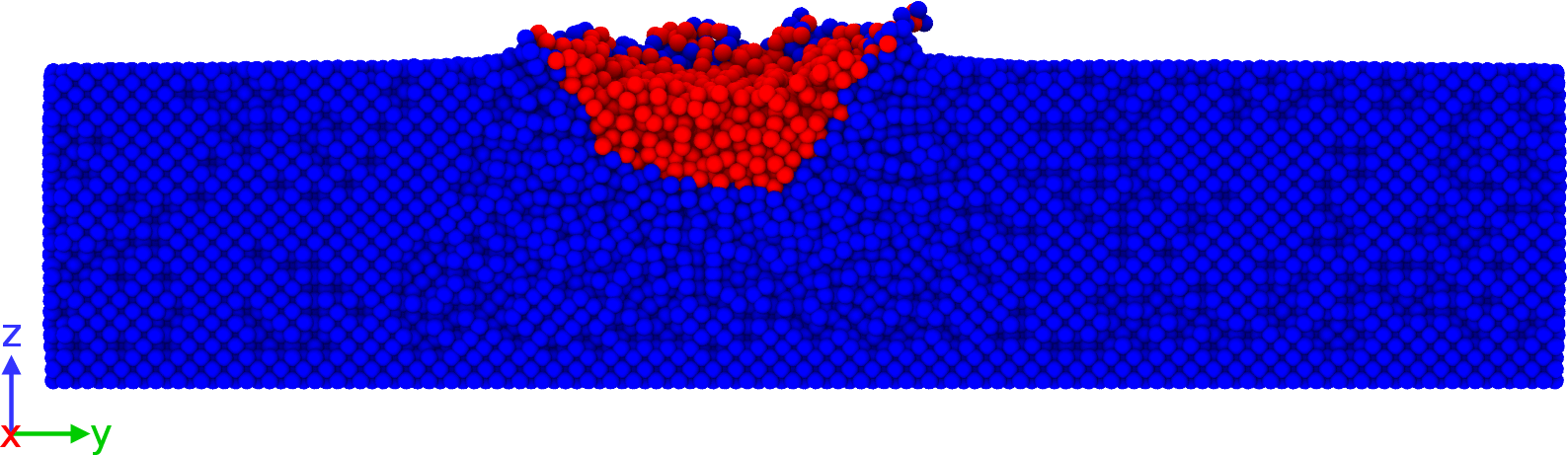}}
	
	\subfigure[$3ps$]{\includegraphics[width=3.9cm]{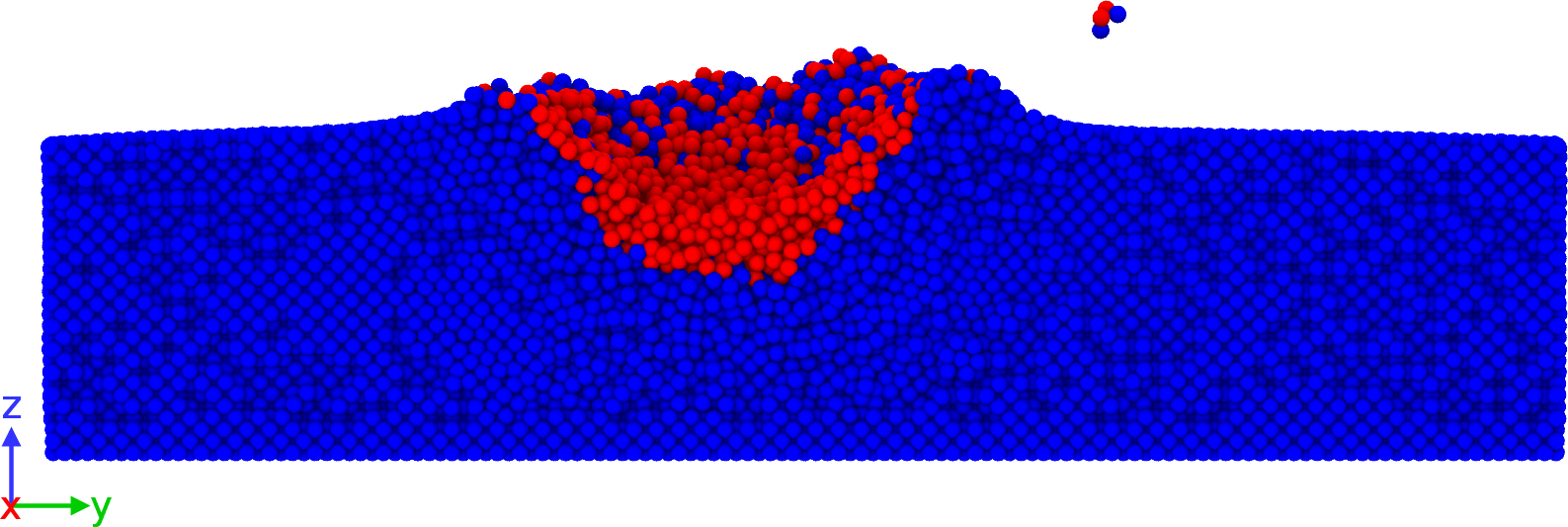}}
	\subfigure[$6ps$]{\includegraphics[width=3.9cm]{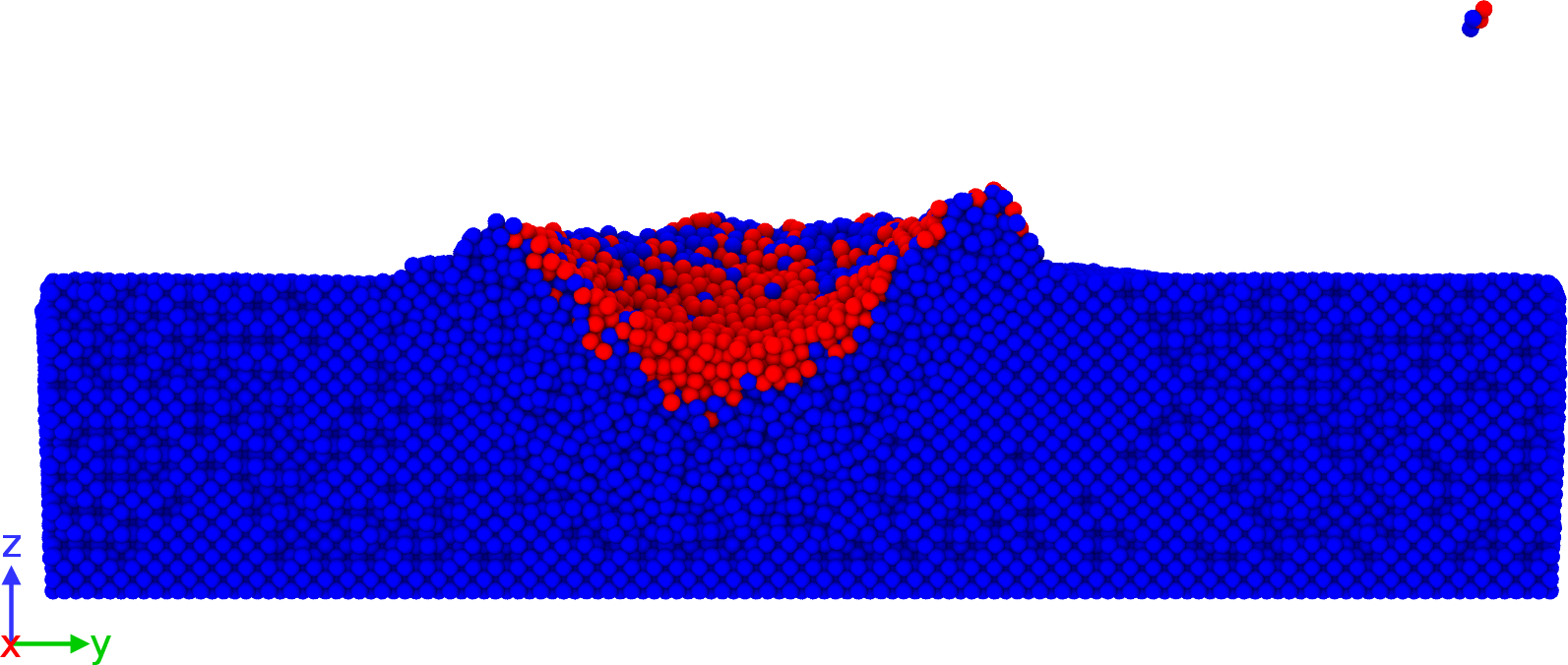}}
	\subfigure[$9ps$]{\includegraphics[width=3.9cm]{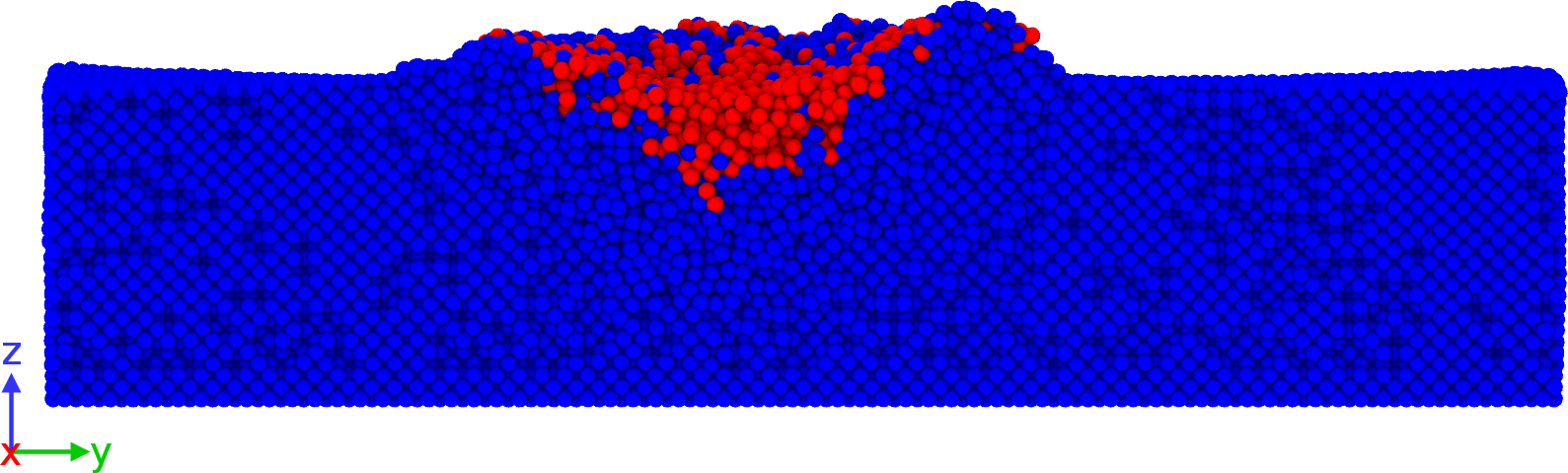}}
	\caption{The MD simulation snapshots for the optimal solution (BPNN-PSO)}
	\label{fig:mdsbp}
\end{figure}

\begin{figure}[htbp]
	\centering
	\subfigure[$1ps$]{\includegraphics[width=3.9cm]{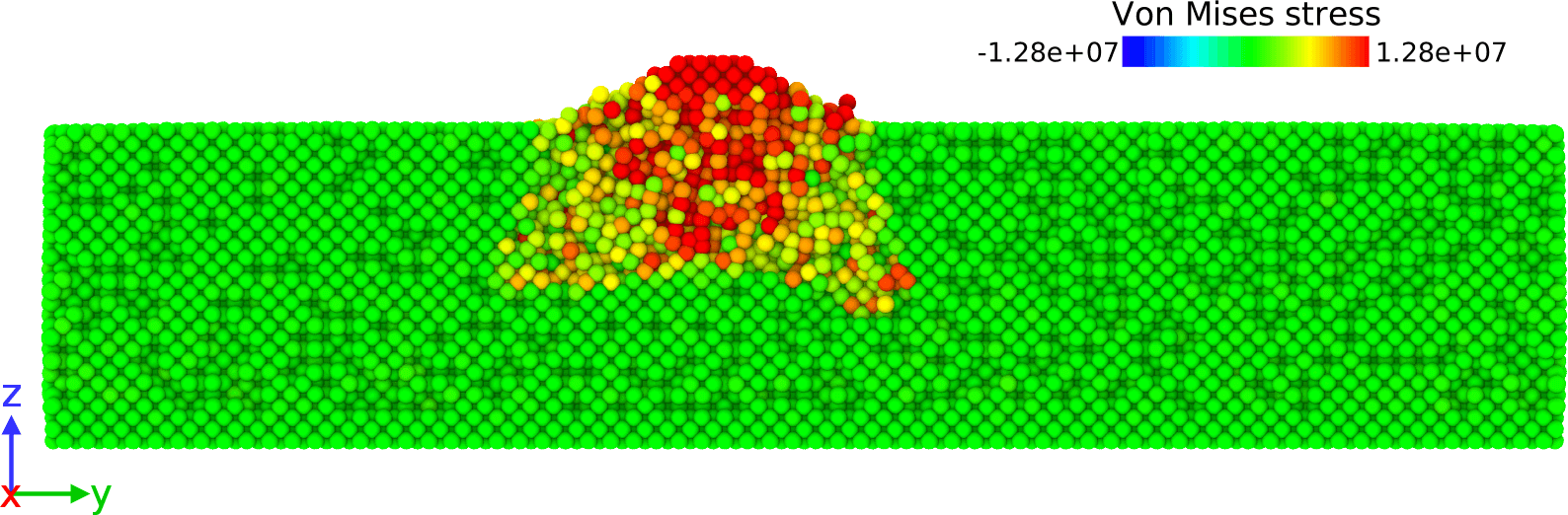}}
	\subfigure[$1.5ps$]{\includegraphics[width=3.9cm]{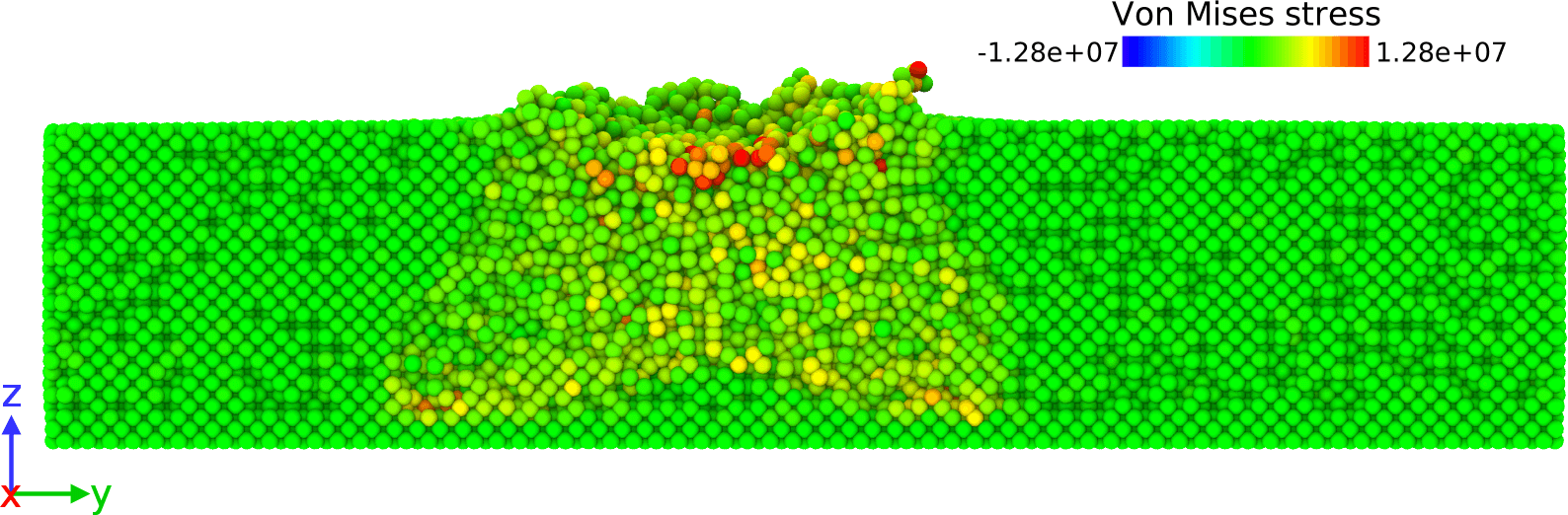}}
	\subfigure[$8ps$]{\includegraphics[width=3.9cm]{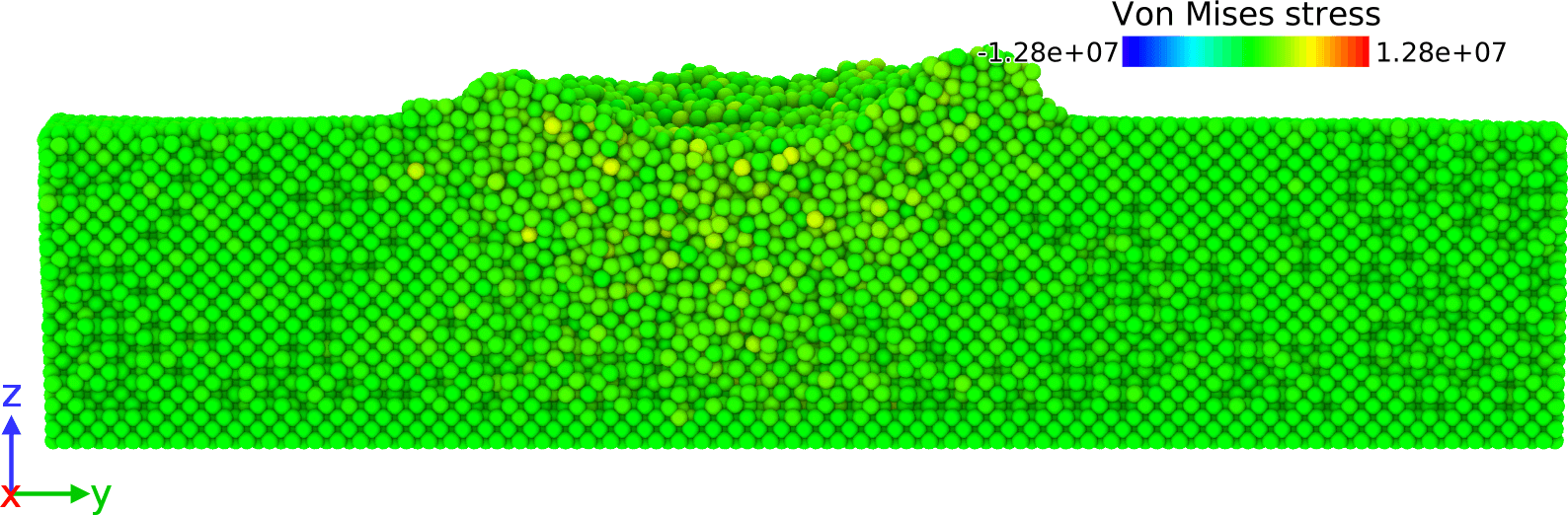}}
	\caption{The Von Mises stress contour plots for the optimal solution (BPNN-PSO)}
	\label{fig:bst}
\end{figure}

As for the results of BPNN assisted optimizations, the convergence curves of the objective function in the optimization procedure is shown in Figure\ref{fig:bpnnopt}. It can be found that both of these methods are available. Moreover, it is obvious that the BPNN assisted PSO method obtained a better solution than other two methods and converged fast. The optimal solution of the above three methods is listed in Table\ref{tab:bpnnsol}, where $1\AA/ps=100m/s$. The comparison of the MD simulation results after impact between BPNN assisted optimizations are shown in Figure\ref{fig:topbpnn}, where the contour plot is according to the Z-coordinate range from $0\AA$ to $10\AA$. Furthermore, the MD simulation snapshots for the optimal solution which obtained by BPNN assisted PSO method are shown in Figure\ref{fig:mdsbp} and the Von Mises stress contour plots are shown in Figure\ref{fig:bst}.

\begin{table}[ht]
	\centering
	\small
	\caption{Optimal solution of BPNN assisted optimization methods}
	\label{tab:bpnnsol}
	\begin{tabular}{c c c c c}
		\toprule
		\multirow{2}{*}{Optimization} &    \multicolumn{3}{c}{Design variables}     & \multirow{2}{*}{Objective function} \\ \cline{2-4}
		                              & $v(\AA/ps)$ & $r(\AA)$ & $\theta(^{\circ})$ &                                     \\ \midrule
		          BPNN-EGO            & 12.000      & 13.1322  & 0.000              & 0.25247                             \\
		          BPNN-PSO            & 12.000      & 15.917   & 6.884              & 0.24658                             \\
		           BPNN-DE            & 12.000      & 15.962   & 6.903              & 0.24661                             \\ \bottomrule
	\end{tabular}
\end{table}

\subsection{Discussions about classic optimizations and BPNN assisted optimizations}

From the above results, it can be found that the performance of BPNN assisted optimizations extremely depends on the accuracy of the BPNN model, so usually the classic optimizations can obtain a better solution than the BPNN assisted optimization methods, but the results of both two kinds of optimization methods are available. Moreover, the computational cost of all the optimization methods is listed as Table\ref{tab:cost}, where $t_p$ means the computational time for just one sample point by MD simulation. Obviously, EGO is the most efficient method for this problem, but PSO and DE obtain a better solution than EGO. Furthermore, it can be found that the BPNN is not suitable for surrogate optimization (EGO) while it did reduce the computational cost of heuristic algorithms (PSO, DE). Besides, another superiority of the BPNN assisted optimization is the portability, it means the BPNN model can be easily applied to any kind of optimization method once the BPNN model has been constructed. Therefore, which kind of optimization method should be utilized is determined by the trade-off between efficiency and accuracy.

\begin{table}[ht]
	\centering
	\small
	\caption{Computational cost of all the optimization methods}
	\label{tab:cost}
	\begin{tabular}{c c c c c}
		\toprule
		Methods  & Modeling cost & Optimization cost & Total cost &  \\ \midrule
		  EGO    & 20$t_p$       & 100$t_p$          & 120$t_p$   &  \\
		  PSO    & -             & 2000$t_p$         & 2000$t_p$  &  \\
		   DE    & -             & 2000$t_p$         & 2000$t_p$  &  \\
		BPNN-EGO & 1000$t_p$     & -                 & 1000$t_p$  &  \\
		BPNN-PSO & 1000$t_p$     & -                 & 1000$t_p$  &  \\
		BPNN-DE  & 1000$t_p$     & -                 & 1000$t_p$  &  \\ \bottomrule
	\end{tabular}
\end{table}

\section{Conclusion}

In this study, a closed loop image aided optimization (CLIAO) method is proposed to improve the quality of deposition during the cold spray process. The main idea of CLIAO method is as follows: firstly, simulate the cold spray process by MD method; then, obtain the value of objective function (flattening ratio) from the result snapshots; finally, find the optimal solution by optimization methods. In order to obtain the flattening ratio from the snapshots directly, the image processing technique is used to generate the required images automatically and calculate the flattening ratio. Moreover, several optimization methods including surrogate optimization (EGO) and heuristic algorithms (PSO, DE) are used to find the optimal solution and the results demonstrated that the PSO method got the best solution while the EGO is the most efficient method. Furthermore, the BPNN is used to accelerate the process of optimization by building a meta-model for objective function and the results shows that the BPNN did improve the efficiency of PSO and DE but it seems not suitable for EGO. In a word, every method has different superiority in efficiency or accuracy, the selection of them should be determined by the trade-off between efficiency and accuracy.

\section*{Acknowledgment}

This work is partially supported by Project of the National Key R\&D Program of China 2017YFB0203701 and the National Natural Science Foundation of China under the Grant Numbers 11572120. This work is also partially supported by the China Scholarship Council.

\section*{References}

\bibliography{mybibfile}

\end{document}